# Recovery of low-magnitude seismic events before the July 29, 2025, Kamchatka megathrust earthquake using waveform cross-correlation enhanced by the addition of stochastic noise

Ivan O. Kitov

## Abstract

The July 29, 2025, Kamchatka megathrust earthquake is one of the largest events in the 21st century. It is likely the best-instrumented case, with numerous stations at regional and teleseismic distances, including high-performance seismic arrays of the International Monitoring System (IMS). The waveform cross-correlation (WCC) method applied to seismic arrays allowed a significant improvement in signal detection accompanied by enhancements in phase association algorithms. Since the WCC detector is the core algorithm of the matched filter method, its efficiency critically depends on the ambient seismic noise coherency with the template signal, with fully stochastic noise providing optimal detection conditions. To suppress the coherent noise component and improve WCC detection, we add a numerically generated random noise sequence adjusted in amplitude to the natural noise. This method is similar to noise "whitening" used in many acoustic applications, but in this study, it is applied before the matched filter is used. The detection threshold gain can be several times higher than that obtained with a pure WCC detector. The recovery of the low-magnitude seismicity obtained by the WCC applied to the IMS arrays and three-component stations prior to the July 29, 2025, Kamchatka earthquake is significantly improved by the random noise addition.

## Introduction

The July 29, 2025, Kamchatka M8.8 earthquake represents an excellent example of observations made by highly sensitive seismic arrays within the International Monitoring System (IMS) of the Provisional Technical Secretariat (PTS) of the Comprehensive Nuclear-Test-Ban Treaty Organization (CTBTO). The original data from the CTBTO observation facilities are available via the virtual Data Exploitation Centre (vDEC) [vDEC, 2026]. The seismic network of the IMS includes 34 operational arrays with apertures ranging from hundreds of meters to tens of kilometres. In its final configuration, the IMS seismic network will comprise 170 stations of two types: arrays and three-component (3-C) stations, to monitor the entire globe. These 170 stations are divided into two major categories: 50 primary and 120 auxiliary stations. The former group is responsible for establishing the statistical significance of detected events. This significance is based on the principal parameters of the associated signals, namely the deviations of travel time, station-event azimuth, and scalar slowness from their theoretical values calculated for signals associated with a given event at primary stations only. The contribution of these parameters to the final event weight, which represents statistical significance, depends on both the station type and the seismic phase. The auxiliary stations assist in refining the estimates of the physical parameters of an event, including its hypocentre, origin time, and magnitude.

The magnitude threshold of the IMS seismic network is not uniformly distributed across the continents, with a legacy emphasis on former nuclear test sites. The Kamchatka region is well covered by IMS array stations, which feature a detection threshold at least four times lower than that of virtually collocated 3-C stations. This provides a dramatic improvement in the magnitude threshold of detected events. The growing number of operational IMS stations over time has led to increasingly better coverage of the Kamchatka region, which now has arrays at almost all azimuths except the southeast. The March 11, 2011, Tohoku earthquake shared approximately the same observation conditions. In contrast, the December 26, 2004, Sumatra earthquake and the largest historical events in the Western Hemisphere lacked such excellent conditions and suffer from higher detection thresholds.

Waveform cross-correlation (WCC) can improve the detection capabilities of array stations by a factor of two to ten, depending on the properties of ambient noise [*e.g.*, Israelsson, 1990; Joswig, 1990; Schaff and Richards, 2004; Gibbons and Ringdal, 2004, 2006; Gibbons *et al.*, 2007; Waldhauser and Schaff, 2007]. This advantage can be used in the study of the earthquake preparation process [Schaff *et al.*, 2025]. The matched filter method is based on WCC, and its performance depends heavily on these noise characteristics [Turin, 1960]. For a matched filter to act as an optimal detector, the ambient noise must be fully stochastic. Under this condition, the absolute value of the cross-correlation coefficient (*CC*) of a template with the noise is minimized on average. Because natural seismic noise consists of regular signals with varying degrees of coherency relative to a given template, the performance of WCC as a detector also varies across a wide range.

The template-coherent component of ambient noise can be suppressed by adding a simulated random noise time series [Press *et al.*, 1986] with either a constant or adaptive amplitude [Adushkin *et al.*, 2025; Dricker and Kitov, 2025; Kitov, 2026c]. The target of this random time series is not the ambient noise as a whole, but specifically its coherent component. When the amplitude of the coherent component is only a fraction of the total noise amplitude, the stochastic time series is highly effective in suppression, bringing the combined noise profile closer to the random state required by the matched filter. Consequently, the sought signal can be resolved even when its amplitude is at just one-tenth of the ambient noise level. However, when the template-coherent component dominates the ambient noise structure—such as during intense aftershock sequences of major earthquakes—the gain achieved by the matched filter with added random noise is substantially lower [Kitov, 2026e].

The ambient noise component not coherent with the template and the sought signal also consists of regular seismic phases. The standard beamforming method is vulnerable to the ambient noise amplitude [Schweitzer *et al.*, 2012]. WCC detection at an array station does not suffer much from these signals [Kitov and Sanina, 2025c]. The difference in the scalar apparent velocities and incident angles of the plane wave of the template and a given regular phase defines the sensitivity of the WCC. The WCC sensitivity to the sought signals is much higher, and the *CC* rapidly decreases with the growing differences in the apparent velocities and incident angles. When the plane waves of regular signals are perpendicular to the plane waves of the templates, these regular signals obtain the properties of stochastic noise and improve the WCC detector performance. When the stochastic noise and regular phases are simultaneously added to the original waveforms, various effects are observed depending on the angle between the regular plane wave simulating noise and the plane wave of the template. These effects depend on the relative amplitude of these two noise sources and on the amplitude of the sought signal [Kitov, 2026e].

In this study, we extend the WCC processing pipeline by adding stochastic noise to the filtered original data in order to suppress the template-coherent component of the ambient noise. Otherwise, the configuration of this study is identical to our previous work [Kitov, 2026d] in terms of geographical boundaries, the time interval, and the set of master events (MEs) along with their associated waveform templates at the same IMS stations. The primary focus is low-magnitude seismicity prior to the July 20 and 29, 2025, Kamchatka earthquakes. The random time series was scaled to a maximum amplitude in a given interval being processed. The average scaling factor (*StN*) was tuned in preliminary calculations [Kitov, 2026e]. Consequently, the WCC detection threshold was reduced by a factor of two, generating a significantly larger number of low-magnitude event hypotheses using the same dataset. The precursory parameters are re-estimated, demonstrating higher statistical significance and greater predictive power.

## Data and methods

### *IMS and IDC data*

The IMS dataset is thoroughly described in [Kitov, 2025d]. The regional and near-teleseismic IMS arrays form the core element of the observational system, with a few regional three-component (3-C) stations assisting in the generation of numerous event hypotheses for the cross-correlation standard event list (XSEL). The raw data were pre-filtered using the same set of five band-pass filters. Not all filters were used at any given station, depending on the local ambient noise characteristics. For example, high-amplitude, low-frequency noise, at the ARCES array (Norway) forces the exclusive use of high-frequency filters. The opposite situation occurs at the CMAR array (Thailand). The filter sets selected for each station remained identical across both studies. There is no difference in the waveform templates used in these two studies, as they are precalculated and remain invariant across all WCC versions.

The weight assigned to a station when determining the statistical significance of an event hypothesis is defined by its historical performance in the Reviewed Event Bulletin (REB) [Coyne *et al*., 2012] of the International Data Centre (IDC). The ratio of events associated with a station to the total number of events in the REB is calculated around each node of a global grid with a spacing of approximately 1.3° [Kitov *et al*., 2016]. The periods before a station was integrated into IDC processing, as well as the periods of downtime for any reason, are excluded from the statistics used to estimate station weights.

The results of WCC processing are compared to the REB. The set of stations in the REB, and the XSEL is identical and a direct phase comparison is possible. This is an important advantage, as internal rules force the IDC to fix many events at the free surface on the assumption that uncertainty in the depth estimates may reduce the efficiency of monitoring under the CTBT mandate [CTBT, 1996]. When an event is fixed at the surface, it cannot be considered 100% natural, since an underground nuclear test cannot be conducted below a 10 km depth [Coyne *et al*., 2012].

For a regular seismicity regime outside the most intensive periods after the largest earthquakes, arrivals in the XSEL and the REB that are close in time (<20 s) can be considered to belong to the same physical event, although their locations can be up to 10° apart. For a quality check, the REB is a reference bulletin to be matched by an automatic XSEL bulletin. The match is defined by at least one common phase, as adopted by the IDC for the quality check of automatic bulletins. The main objective of WCC processing is to find XSEL events not matching any REB event—that is, to create the REB-ready event hypotheses in addition to the REB [Bobrov *et al*., 2014; Bobrov *et al*., 2016a,b; Bobrov *et al*., 2017; Kitov and Sanina, 2025a,b]. The statistical significance of these new events depends on the capability of the XSEL to successfully match almost all REB events in a specific area [Kitov, 2026a,b,d].

### *Variation of seismicity within the Kamchatka region*

The earthquake preparation process is localized relative to global seismicity, yet it remains intrinsically dependent on the evolution of global tectonics. The boundary conditions for a given megathrust earthquake are difficult to delineate precisely because they evolve continuously alongside conditions in the surrounding regions. Conversely, any megathrust earthquake fundamentally alters the stress-strain state far beyond its immediate aftershock zone. While the largest earthquakes are typically repeatable in terms of their locations and focal mechanisms, the characteristics of local seismicity often diverge from those of the broader seismic regions as delineated by the Flinn–Engdahl (F-E) regionalization scheme [Flinn and Engdahl, 1965]. The hypocentre of the July 29, 2025, Kamchatka earthquake (J29) is located within F-E seismic region 19 and is directly associated with the subduction of the Pacific plate. On a broader scale, seismicity patterns to the southwest and northeast of this location (shown in Figure 1) exhibit distinct differences. Specifically, seismicity within the narrow epicentral zone of the J29 earthquake—centred around 52.0°N, 160.0°E—is characterized by extremely dense seismicity. This zone extends approximately 150 km in length and 60 km in width. The high density of seismic events reported by the IDC since 2001, along with the locations of several major earthquakes (including the J29 event), is directly correlated with the high cumulative seismic energy released from this spot.

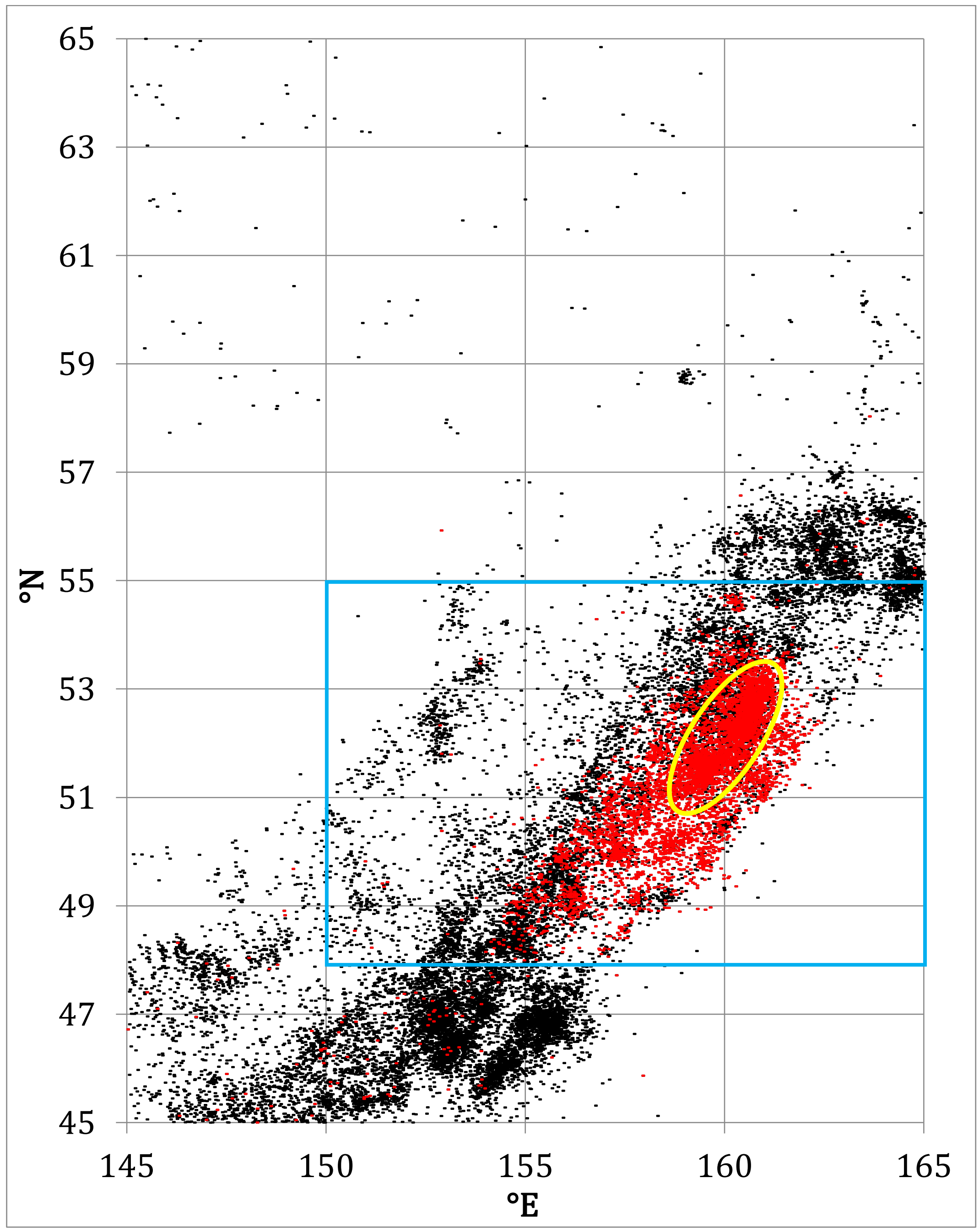

Figure 1. Earthquakes in the studied area between January 2001 and December 2025 as per the REB. Events after July 20, 2025, are shown in red. The blue rectangle indicates the aftershock zone, and the yellow oval represents the highest seismicity spot

Seismic energy represents only a small fraction of the total elastic strain energy accumulated prior to its release during the observed seismic process. As the concentration of elastic energy approaches a critical threshold, the earthquake preparation process is accompanied by an increasing number of low-magnitude events, characterized by both an increasing event count and a greater total seismic energy release. A monitoring system must focus on regions of steadily increasing seismicity and the specific zone of the impending rupture. The latter may also reside in a pre-critical state, though likely with some delay, as suggested by the failure of the July 20, 2025, earthquake rupture to propagate through what would become the July 29 rupture zone.

The local conditions surrounding the epicentre of the J29 event differ markedly from those of the adjacent areas. Thus, the specific earthquake preparation process reflected in the low-magnitude seismicity may diverge in several ways, including the characteristics of the recurrence curve. This curve scales the frequency distribution of detected earthquakes according to their magnitudes and is defined by a site-specific slope (*b*-value) and a corner magnitude. These two parameters are useful for predicting the magnitude distribution of missed events below the completeness threshold. The WCC method enables the recovery of numerous earthquakes omitted from standard catalogues. The frequency distribution of these newly detected events is expected to follow the linear extrapolation of the standard recurrence curve below the completeness magnitude [Gutenberg and Richter, 1954; Scholz, 1968]. The blue rectangle in Figure 1 bounds the J29 aftershock zone, which includes the cluster of the highest seismic activity. XSEL events within this zone, generated by our modified WCC pipeline, are analyzed to identify features potentially linked to variations in the seismic process parameters prior to the July 20 (J20) and July 29, 2025, earthquakes.

Figure 2 compares the recurrence curves for the studied earthquake preparation zone defined by the blue rectangle ("ASPERITY", A-zone) to those for the areas surrounding this zone ("OUTASPERITY", O-zone) and illustrates the difference in the slopes of the curves obtained for shallow (depth <80 km) seismicity. For deep events, there is no tangible difference in the slopes between these two zones. The extension of the recurrence curves into the low-magnitude range below the corresponding corner magnitudes allowed for estimation of the number of events missing from the REB in a given $m_b$ interval. The lower the $m_b$ value is below the corner magnitude where the WCC can find events, the larger the number of these missed events.

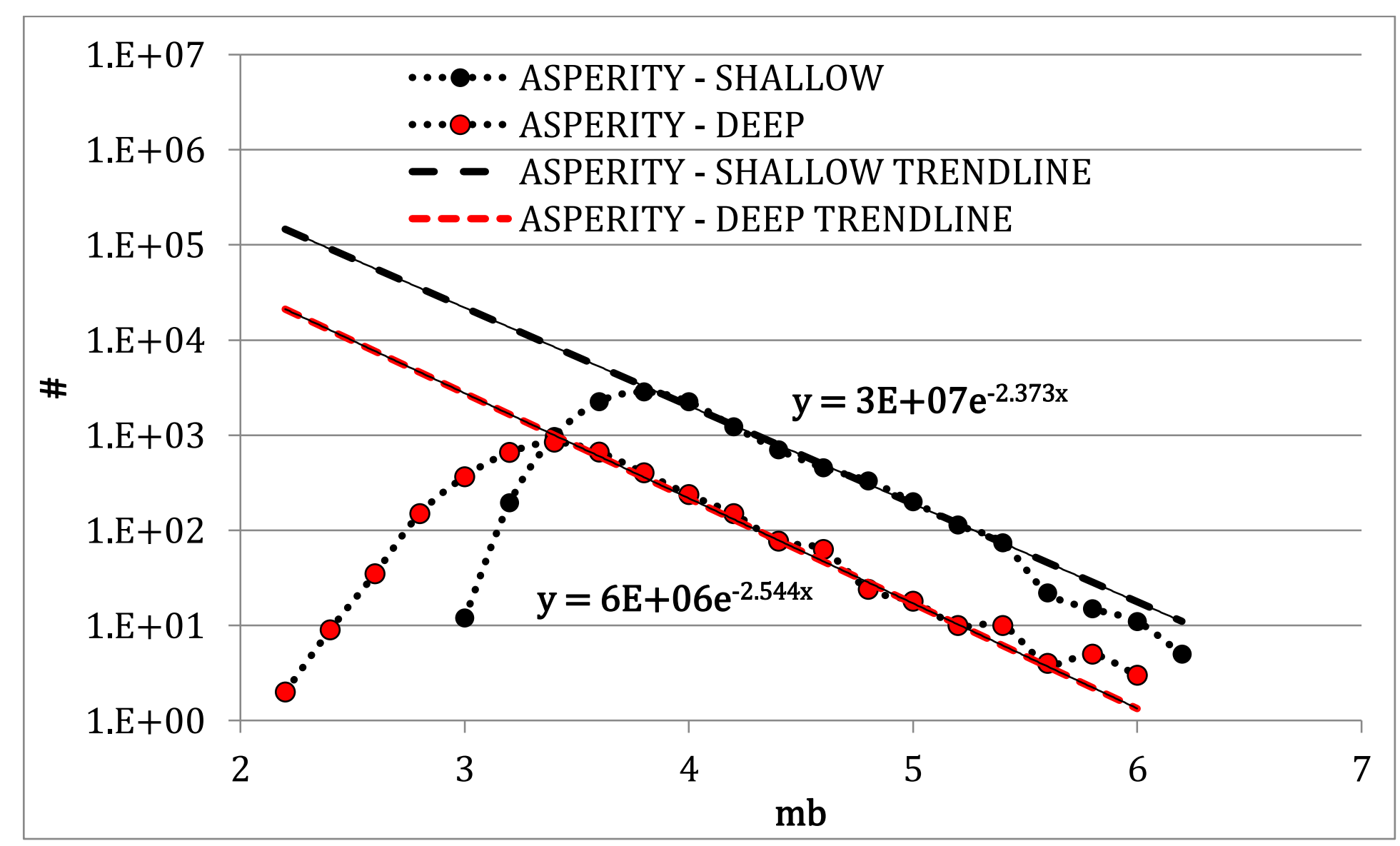

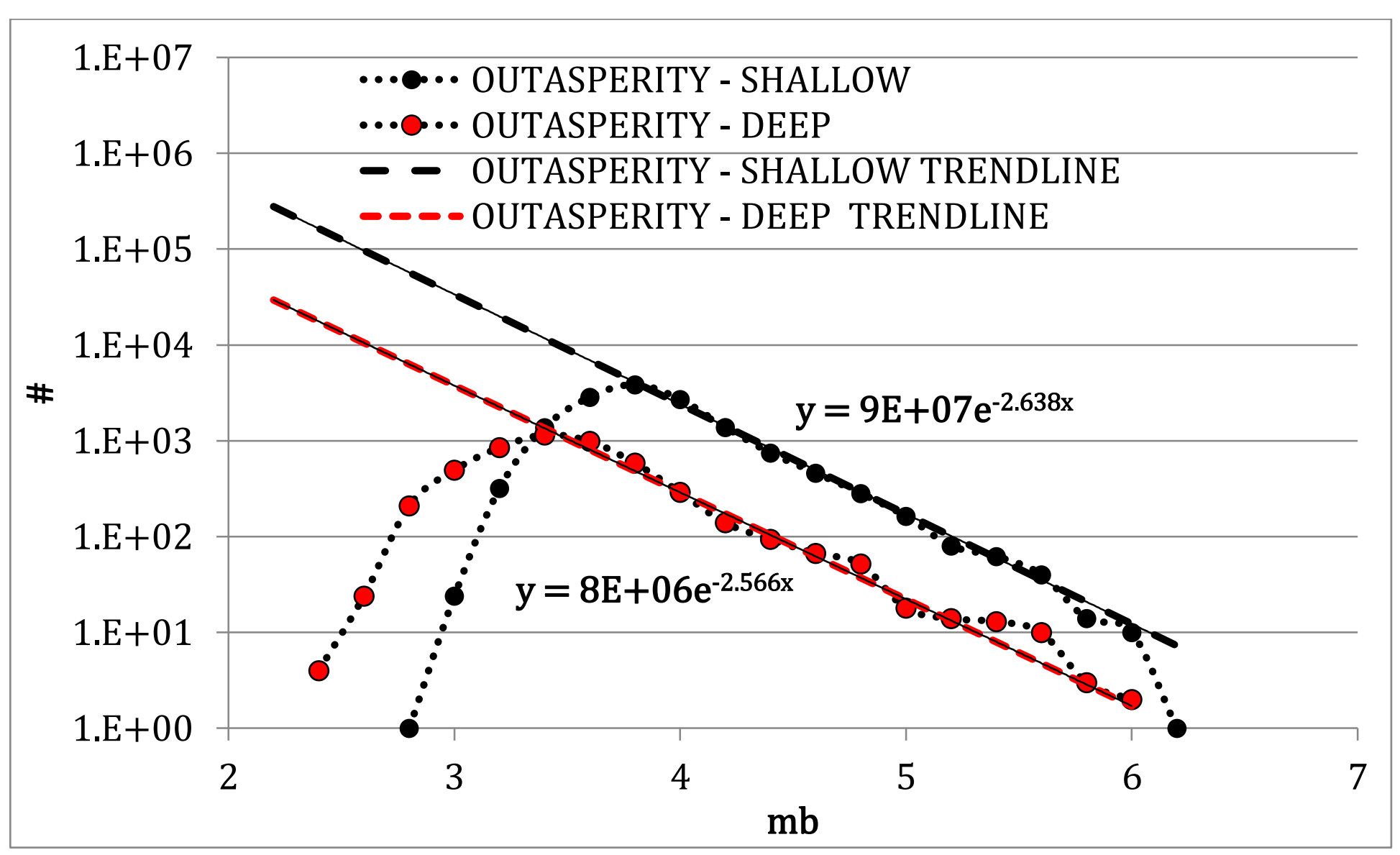

Figure 2. Recurrence curves for the "ASPERITY" and "OUTASPERITY" zones. Shallow and deep (depth > 80 km) events are shown separately. The trend lines for shallow earthquakes are characterized by different slopes in the two zones, likely reflecting significant variations in seismic regimes. For deep events, the difference between the trend line slopes is negligible.

The seismic process within the whole studied region (45°N–65°N, 145°E–165°E) reported by the IDC for the period between July 2 and August 3, 2025, is illustrated in Figure 3. Body-wave magnitudes of the IDC-detected earthquakes are shown as a function of time. The events with magnitudes around 5.5 occurred outside of the July 20, 2025, M7.4 earthquake and prior to the J29 event are highlighted in red. These events can potentially disturb the preparation process of these two largest earthquakes. The properties of the seismic process in Figure 3 prior to the July 20 (M7.4) and 29 (M8.8), 2025, earthquakes do not demonstrate any specific features indicating the approaching megathrust event. The J20 event can be tested as a J29 foreshock, but a few similar events in the past were not followed by M>8.5 earthquakes. The J20 is of interest in itself. The preparation process of an M7.4 event is worth investigating in detail as the events accompanying the evolution of cracking and faulting leading to the rupture can be recovered by using the WCC pipeline.

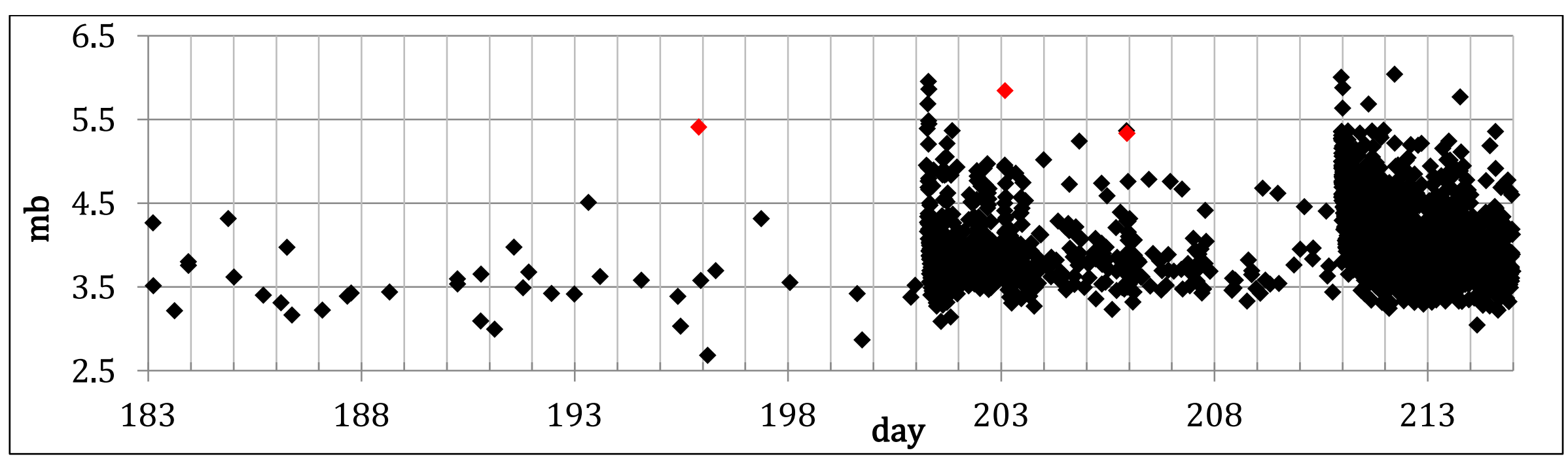

Figure 3. The earthquakes reported by the IDC between 2025193 and 2025212 within the studied regions in Figure 1. No REB events were detected on 2025189.

The resolution of the WCC can be adjusted to the parameters of seismicity to be recovered below the completeness threshold in a relatively wide range of magnitudes. The original study [Kitov, 2026d] used just a pure WCC approach and allowed for the detection of numerous statistically significant XSEL events starting from the corner magnitude and below. In this study, the WCC resolution is enhanced by the stochastic noise addition, leading to the increasing values of SNRcc for many of the detections obtained by the pure WCC version and to the detection of new sought signals. Also, some potentially irrelevant detections in the pure WCC version are suppressed and disappear from the enhanced version.

### *The WCC pipeline*

A specialized version of WCC processing is used to recover the evolution of low-magnitude seismicity before a major earthquake. A set of detection thresholds was tuned to generate 30 to 120 detections per hour at IMS arrays and 3-C stations associated with the REB events in the studied part of the Kamchatka region (see Figure 1). The first P-wave signals used as waveform templates for cross-correlation were selected based on the properties of the REB events they were associated with. At the global level, the best events were selected in a thorough cross-correlation exercise with all pairs of REB events around the nodes of a global grid with an almost constant spacing [Kitov *et al.*, 2016]. The REB events with the largest number of detected neighbours were promoted to the final list of approximately 40,000 MEs used in a prototype WCC pipeline tested at the IDC [Bobrov *et al.*, 2014; Bobrov *et al.*, 2016a,b]. This pipeline includes three major stages: detection, phase association, and conflict resolution.

The WCC detection follows the matched filter method and seeks to maximize the signal-to-noise ratio (SNRcc) in the trace of the cross-correlation coefficient (*CC*). The detection procedure is flexible and includes a set of five band-pass filters, cross-correlation window lengths adjusted to the properties of the sought signal, and adaptive detection thresholds to provide a predefined detection rate. Standard SNR values and RMS amplitudes in the winning filter-cross-correlation window pair is also calculated for each WCC detection. The WCC detection procedure generates a list of arrivals of the sought signals with all defining parameters used in the subsequent Local Association (LA) and Conflict Resolution (CR) procedures [Kitov, 2026b,d].

For a given ME, the detected signals associated with the final list of XSEL event hypotheses are considered valid. Unassociated detections can be generated by actual physical sources close to the MEs—that is, the sought events, but not supported by detections from the same source at different stations. They are considered conditionally false, since they can be converted into valid detections depending on the parameters of the LA and CR. As an alternative, some detections can actually be false, as they are generated by events far away from the ME. To distinguish between valid and actually false detections, the performance of the WCC pipeline with actual data is compared to a case where the data are represented by numerically generated random noise [Kitov, 2026e].

The LA process represents a grid search for potential events in the vicinity of an ME. The search radius and the grid spacing are adjusted to the slowness of the template and the sought signals. P-phases with low apparent velocities need a smaller grid, whereas for teleseismic P-phases the grid has to be larger. The distance between nodes can compensate for the difference in the search radii. The Event Definition Criteria (EDC) must provide the final XSEL with high statistical significance compared with those generated by a stochastic time series. The principal rule of three or more stations being associated with a valid event is borrowed from the IDC [Coyne *et al.*, 2012].

The EDC is flexible and has to be adjusted to the case. For example, regular seismicity in the studied region observed before the J20 event is characterized by a few REB events per day. There are several three-day-long periods without events. By contrast, there are more than seven hundred aftershocks in the REB on July 30, 2025. There were no events with magnitudes below 4.5 during the first hour after the J29 event [Kitov *et al.*, 2026]. This dramatic difference requires adjustments to the EDC in order to avoid numerous phase misinterpretations and misassociations. For background seismicity, the target is events with lower magnitudes, and the EDC have to be adjusted accordingly. The earthquake preparation process involves events far below the corner magnitude in Figure 3, and thus, the thresholds of the parameters in the EDC have to be lowered to levels germane to the task. This is the approach to defining the set of minimum requirements for a valid event adopted by seismological agencies. Such a set has to guarantee at least the minimum level of statistical significance for the events in the final bulletins. Human review is usually a mandatory step to confirm that the associated signals are appropriate and reliable. Both automatic and interactive bulletins follow the predefined EDC, which have to be tuned via trial-and-error procedures [Saragiotis and Kitov, 2020]. At the IDC, a five-year-long GSETT-3 exercise was accomplished [Ringdal, 1994] to tune the current EDCs to the analyst workload.

When investigating low-magnitude events below the visibility of experienced analysts, one has to rely on results obtained with various EDC sets in comparison with a fully stochastic waveform on one side and to reliable REB events matched by XSEL events on the other side. Without a priori information, twelve different sets of EDCs were introduced in an attempt to cover the magnitude range below the completeness threshold of the recurrence curve for the studied Kamchatka region [Kitov, 2026d]. There were two basic versions of the LA procedure: strict and weak. The strict LA covers the range closer to the corner magnitude and the transition from new XSEL events to those matching the REB. All thresholds of the EDC defining parameters are listed in [Kitov, 2026d]. The weak LA version is aimed at the lowest magnitudes and has lower thresholds for the EDC-defining parameters. In order to obtain various corner magnitudes in the recurrence curve based on the XSEL events, a set of six origin time tolerances is used for the strict and weak LA versions: 5.0, 3.0, 2.0, 1.0, 0.5, and 0.25 seconds. Altogether, there are twelve different EDC sets which are also used in the current study with enhanced WCC detection capabilities. The comparison of the original XSEL and the improved XSEL versions is straightforward—through the numbers and distribution of the XSEL events.

The CR procedure is an important part of WCC processing. It has not been changed for the enhanced WCC version and follows the same rules. When two event hypotheses generated by two different MEs compete for the same detection (within 20 s), the weights of the hypotheses are compared first. If the weights are equal, the number of associated stations defines the winner. If both parameters are equal, the RMS origin time is used to resolve the conflict.

***Master events***

There were 100 MEs used in the previous study with the WCC processing without random noise added to the original waveforms. The same MEs are used in this study. Figure 4 presents the distribution of these MEs with 63 MEs in the “ASPERITY” zone (A-zone) marked in yellow. The remaining 37 MEs are in the “OUTASPERITY” zone (O-zone). They are needed to reduce the effect of the side sensitivity of the A-zone MEs to the sought events located in the O-zone [Kitov, 2026e]. The XSEL events have to be located in the A-zone to be counted in the precursory parameters. The MEs in the A-zone can create XSEL events in the O-zone and vice-versa because the grids in the Local Association process has radii of 48 km and 81 km for the strict and weak LA settings, respectively [Kitov, 2026d]. Therefore, the XSEL events in the A-zone are a more reliable estimate of the actual seismicity than the events created by the A-zone MEs. However, the difference between these two counts is not significant.

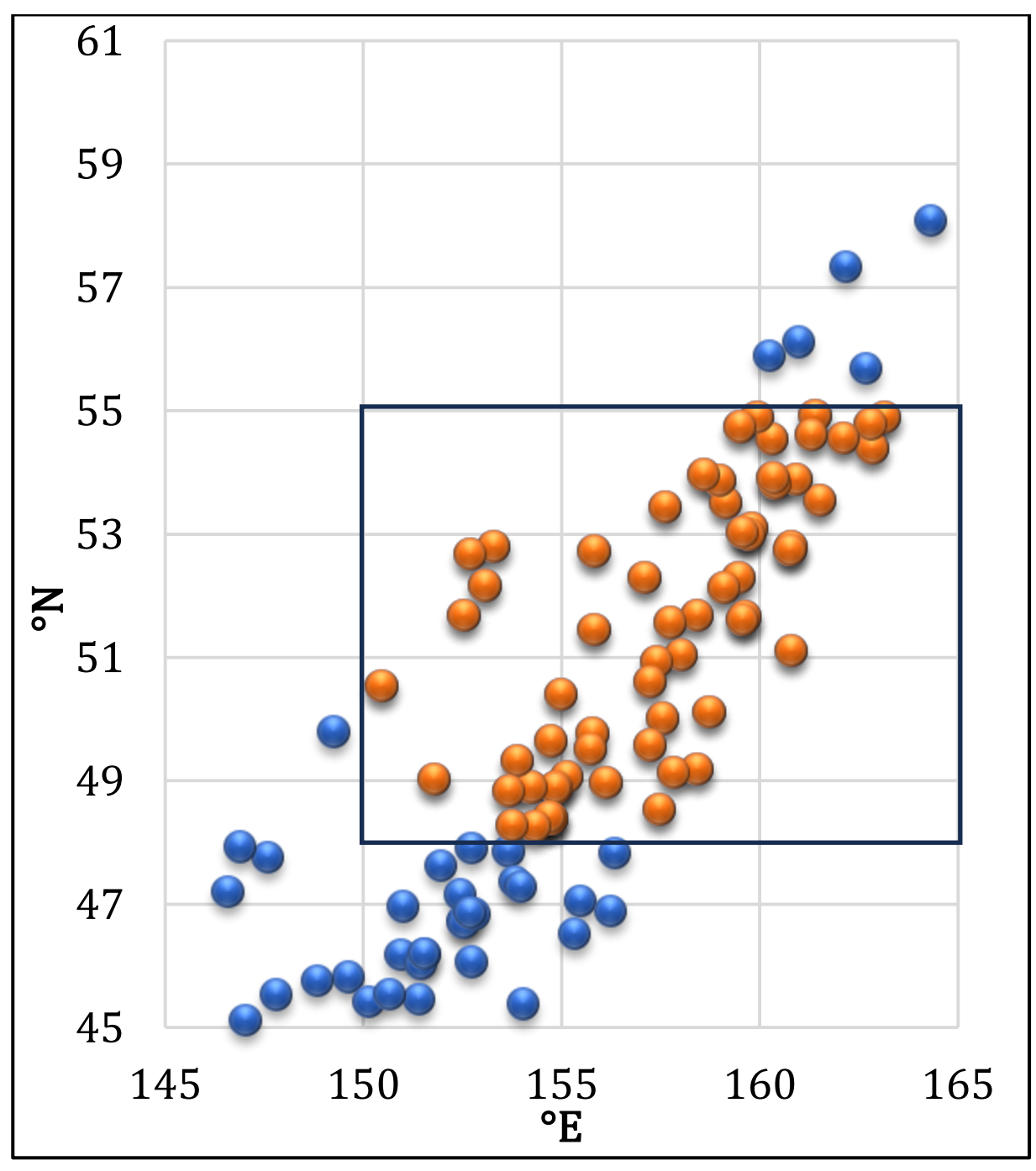

Figure 4. Sixty-three master events within (red circles) and thirty-seven outside the ASPERITY zone (blue circles) – the presumed earthquake preparation area 45°N-55°N, 150°E-165°E.

The distribution of the MEs was approximately even across the whole studied region and covered the depth range from 0 km to 700 km. The REB events serving as MEs were selected by a WCC procedure that provides the best performance of a given ME in finding its neighbours from the REB. These neighbours do not necessarily require accurate locations as the IDC applies the fixed depth rule to all events with high depth uncertainty which shifts their hypocentres. The smaller an event is, the larger the shift can be. Therefore, not all REB events can be found by the WCC, whose performance critically depends on the physical distance between the master and sought events. There is a portion of the REB events not matched in the XSEL because they are not physically close to the MEs although their hypocentre estimates are.

### *Optimizing the WCC detection process with added stochastic noise*

The effect of the computer-simulated stochastic noise on the overall performance of the WCC pipeline was demonstrated in [Kitov *et al.*, 2026; Kitov, 2026a]. Stochastic noise with the amplitude scaled to the peak waveform amplitude in a given time interval using a *StN* factor of 10 to 1000 completely suppressed the possibility of detecting the sought signal using a given template, even if the sought signal was identical to the template [Kitov, 2026e]. The result of the WCC processing with *StN*>10 was fully random and the created event hypotheses were used as a basis for the estimation of the statistical significance of the original WCC event hypotheses without the added stochastic noise.

To understand the transition from the original state to fully stochastic waveforms, the *StN* factor was progressively increased in the range from 0 to 1000 with a few small steps at the beginning: 0.005, 0.01, 0.05, 0.1, 0.5, and 1.0. A one-day-long period prior to the July 29, 2025, Kamchatka earthquake was selected for a full WCC processing within the region with 20 master events (MEs) evenly distributed across the studied region. Figure 5 presents the results of the WCC processing with twelve configurations of the LA parameters in the pipeline using a unique WCC detection list [Kitov, 2026e]. These configurations are indexed from 1 to 12. The weak version with 5.0 s origin time tolerance has index 1 and the strict version with the 0.25 s tolerance has index 12. Decreasing the tolerance from 5.0 s to 0.25 s for a given LA version provides a smooth increase in the corner magnitude of the recurrence curve obtained from the XSEL [Kitov, 2026d].

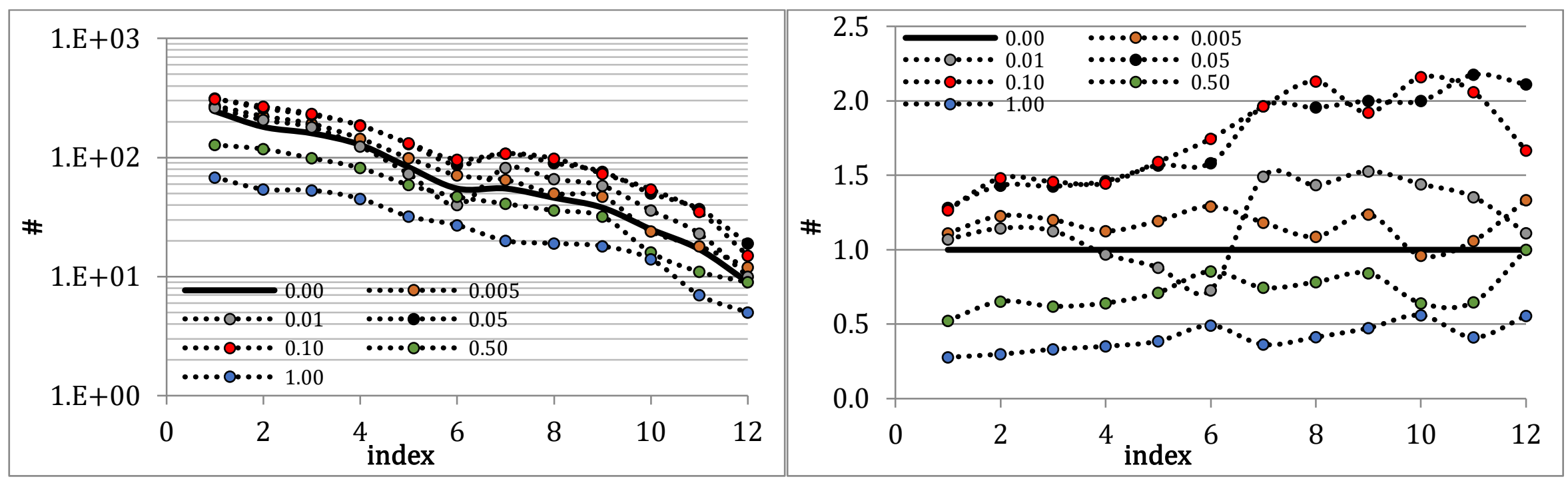

Figure 5. Variation of the number of XSEL events as a function of *StN*. a) The event counts for number for seven *StN* values from 0.0 to 0.50 for the 12 LA version-origin time tolerance cases. b) Ratios of the non-zero *StN* curves to the baseline curve with *StN*=0.

The ratios of the nonzero *StN* curves to the original one with *StN*=0 demonstrate a gain in the number of XSEL events by up to a factor of two for larger indices at the *StN* values of 0.05 and 0.1. From Figure 5, we estimated an optimal *StN* value of 0.075 for the WCC processing. This specific value is used here to improve the earthquake prediction parameters introduced in [Kitov, 2026b] and further developed in [Kitov, 2026d]. The main objective is the July 29, 2025, Kamchatka earthquake, while the July 20 event is also of interest.

In order to maximize the detection sensitivity and resolution, a single value of 0.075 is likely not enough. The template-coherent component in the ambient noise may vary in amplitude within the studied interval, and these fluctuations are not necessarily synchronized with the other components in the ambient noise. The maximum amplitude in one-hour-long processing windows can vary across a wide range, independently of the seismicity in the studied region. Then, the *StN* values would be inappropriate for the case and negatively affect the WCC performance. This influence is mainly local to the processing interval and does not spread to the whole processing period between July 6 and July 31, 2025. The current feasibility study primarily evaluates this approach to *StN* application. Even one *StN* value for the whole period can significantly improve the XSEL in both number and quality.

For the maximum amplitude *StN* scaling, several values can also be selected within a certain range, with the optimal one from the full set providing the largest SNRcc for a given detection. However, this approach requires a manifold increase in computational resources. There are local measures of amplitude such as STA and LTA, which must be tested. The detection rate does not change much with the addition of stochastic noise, as the detection threshold is adjusted to the predefined detection rate range.

## Results

### *Performance difference with and without stochastic noise component*

We have processed the period between July 6 (2025187) and July 31 (2025212), 2025, using the *StN* value of 0.075 and the same WCC pipeline as in the previous study without stochastic noise [Kitov, 2026d]. The resulting XSEL bulletins calculated for the six-hour-long intervals were smoothed by a moving sum of four quarter-days (MS(4)). Figure 6 presents the MS(4) curves for the weak and strict LA versions separately. Only the XSEL events within the A-zone were counted as potentially related to the preparation of the J29 and J20 earthquakes.

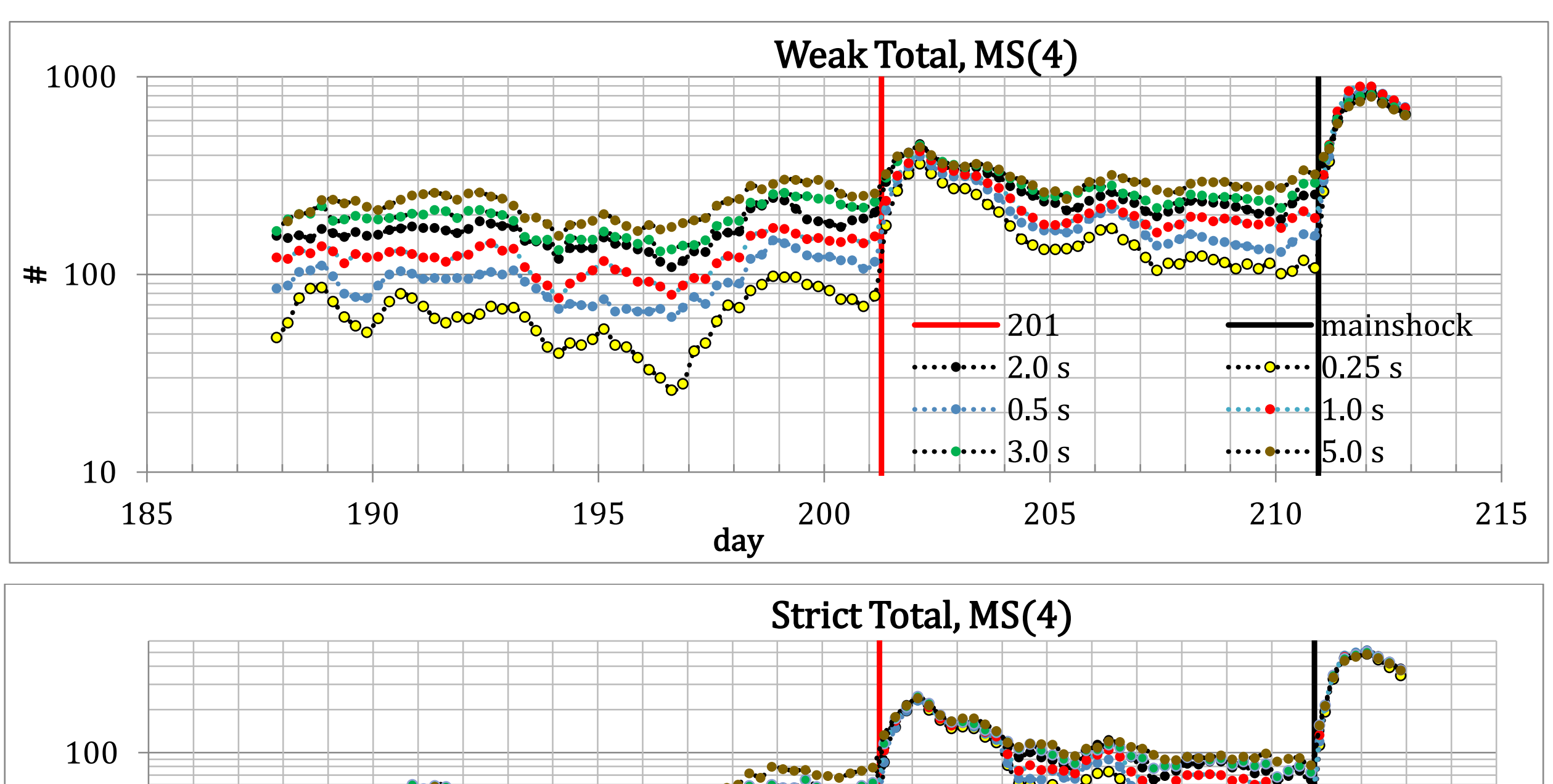

Figure 6. The evolution of the daily number of XSEL events with a six-hour step. The upper panel shows six tolerance curves for the weak LA version. The lower panel shows the same for the strict LA version. The times of the J29 event (mainshock) and J20 event (201) are indicated by vertical lines.

The evolution of seismic activity shown in Figure 6 has three distinct periods: 187-201, 201-210, and 211-212 in accordance with the influence of the J20 and J29 events. Before the J20 earthquake, the level of activity was relatively low, but the number of XSEL events for any of the twelve indices and their corresponding EDCs was much larger than that reported in the REB and shown in Figure 3. For the weak LA version, there were tens of events per day even for the most conservative origin time tolerance of 0.25 s. For the strict LA version, the number was lower, but still above the number of events reported in the REB. This is the effect of the WCC superiority in sensitivity and resolution, also enhanced by the addition of the case-tuned random noise. All curves demonstrate some positive trends after day 2025196 toward the J20 event. These trends have peaks between 190-200 and many of them fall prior to the J20 earthquake. Many curves demonstrate a quick rebound just before the J20 and J29 events.

The peak of seismic activity right after the J20 event smoothly transformed into a slowly decaying trend characterized by the divergence of the curves with different tolerances. Near the peak, all curves converge, illustrating the absence of low-magnitude events detected amid the intense seismic noise generated by the J20 M7.4 earthquake and its largest aftershocks. This noise is highly correlated with the sought signals, and thus detection of smaller events is fully suppressed. The same effect is observed after the J29 earthquake.

The differences between the weak and strict curves for the same origin time tolerance indicate the change in seismic activity in the magnitude interval defined by the LA parameters for two versions, although the magnitude boundaries of these intervals are not known. One can test the twelve indices as a series of potential corner magnitudes of a generic recurrence curve as described in [Kitov, 2026d]. Figure 7 illustrates such recurrence curves obtained for three different periods of seismicity: the days before J20 events (187-201), the days with intense aftershock sequence between the J20 and J29 events (201-210), and the whole period between July 6 and July 31. Only the XSEL events within the A-zone were counted.

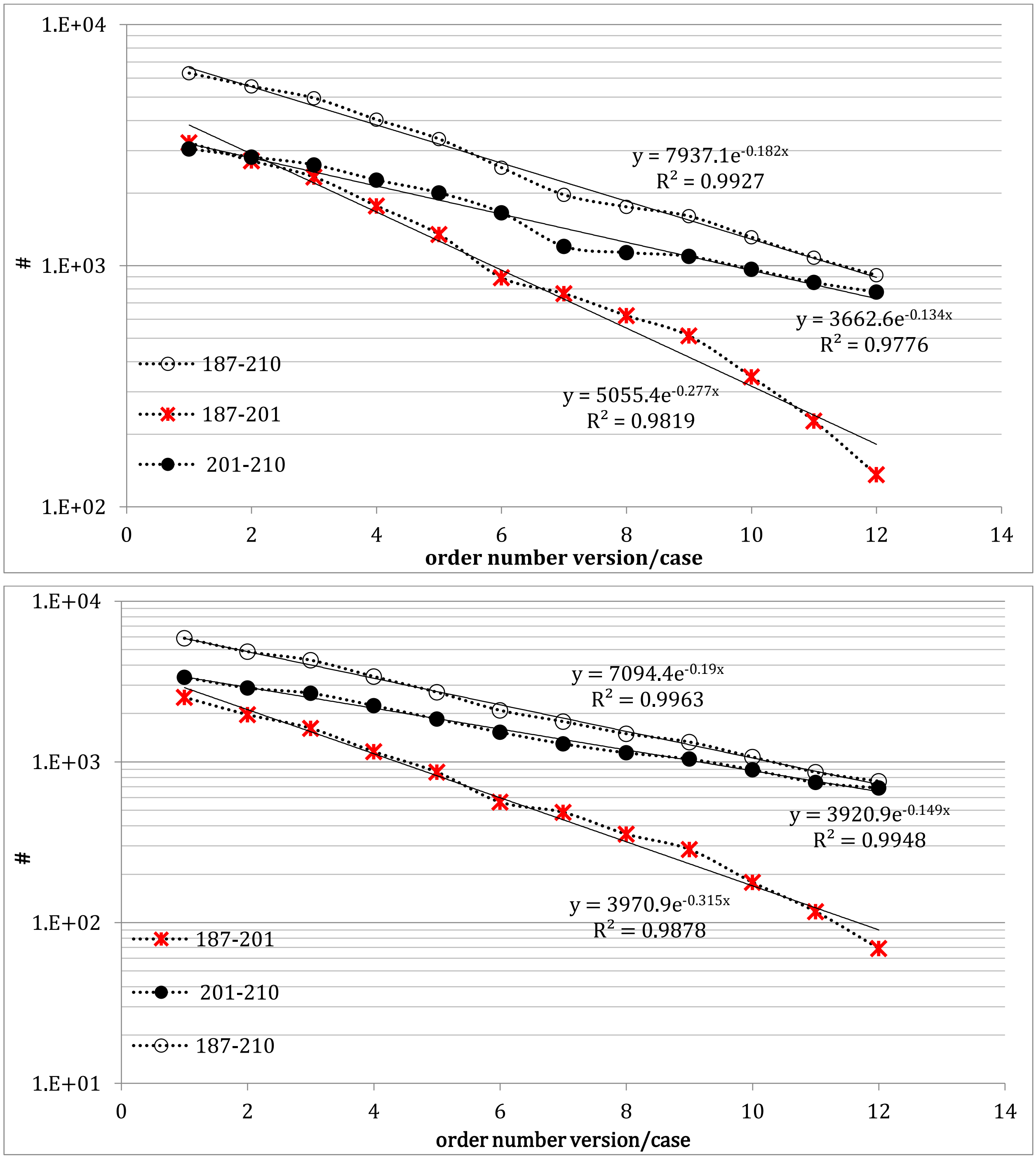

Figure 7. Recurrence curves for the XSEL events within the A-zone produced by the 100 best MEs in various time intervals. Upper panel: *StN*=0.075. Lower panel: *StN*=0.0.

The upper panel illustrates the result obtained in the current study. The curves for the first and second time periods have different slopes and levels expressing the difference between a regular seismic regime and the aftershock sequence. The regression lines for these two curves have $R^2$~0.98, which is very close to standard recurrence curves obtained from the REB or any other catalogue based on body-wave magnitude estimates. In the lower panel, similar curves for the XSELs in the A-zone are presented for the case *StN*=0.0 using the results described in the previous study [Kitov, 2026d].

Comparison of the *StN*=0.075 and 0.0 cases is important for the understanding of the magnitude range covered by the twelve magnitude thresholds defined by their respective EDCs. The *StN*=0.075 curves for the 201–210 time interval and for the whole period do not intersect as the input of the interval prior to the J20 event is substantial. This input is significant for all twelve EDCs meaning that the index 12 has the magnitude threshold below that of the REB as the XSEL for the 201-210 time interval includes almost all the REB events. For the *StN*=0.0 case, these two lines intersect exactly at the index 12 indicating the magnitude thresholds of the $EDC_{12}$ and the REB are very close. The overall curve for *StN*=0.0 demonstrates an exponential distribution ($R^2$=0.99) of the number of XSEL events for the twelve indices, which are just a sequence of integer numbers. This observation makes it possible to convert the indices into a sequence of magnitudes with a constant step using a simple linear relationship [Kitov, 2026d].

In the *StN*=0.0 case, the $EDC_{12}$ was designed to have the completeness magnitude close to the corner magnitude of the REB curve—that is, to generate approximately the same number of XSEL events as in the REB for periods of low seismicity. For the *StN*=0.075, the same $EDC_{12}$ undergo a shift toward the events with lower magnitudes as the SNRcc values for the detections increase relative to the *StN*=0.0 case and many new sought signals are detected above the SNRcc threshold. To illustrate this feature, Figure 8 presents the MS(4) curves for the *StN*=0.0 and 0.075 cases having similar numbers in the respective XSELs. For the weak version in the upper panel, the *StN*=0.075 curve with 0.25 s tolerance is above the *StN*=0.0 curve with 0.5 s tolerance except the periods after the J20 event.

A more important effect is the shift observed in the strict LA version curves shown in the lower panel. The *StN*=0.075 curve with 0.25 s tolerance is above the *StN*=0.0 curves with 0.5 s and compatible with the curve with 1.0 s tolerance during periods before the J20 and J29 earthquakes. During the high seismic activity after these earthquakes, the effect of *StN*=0.075 is negative – too many event hypotheses lead to mutual annihilation, leading to a fall in the XSEL numbers. The stochastic noise enhancement has to be focused on the periods of regular seismicity.

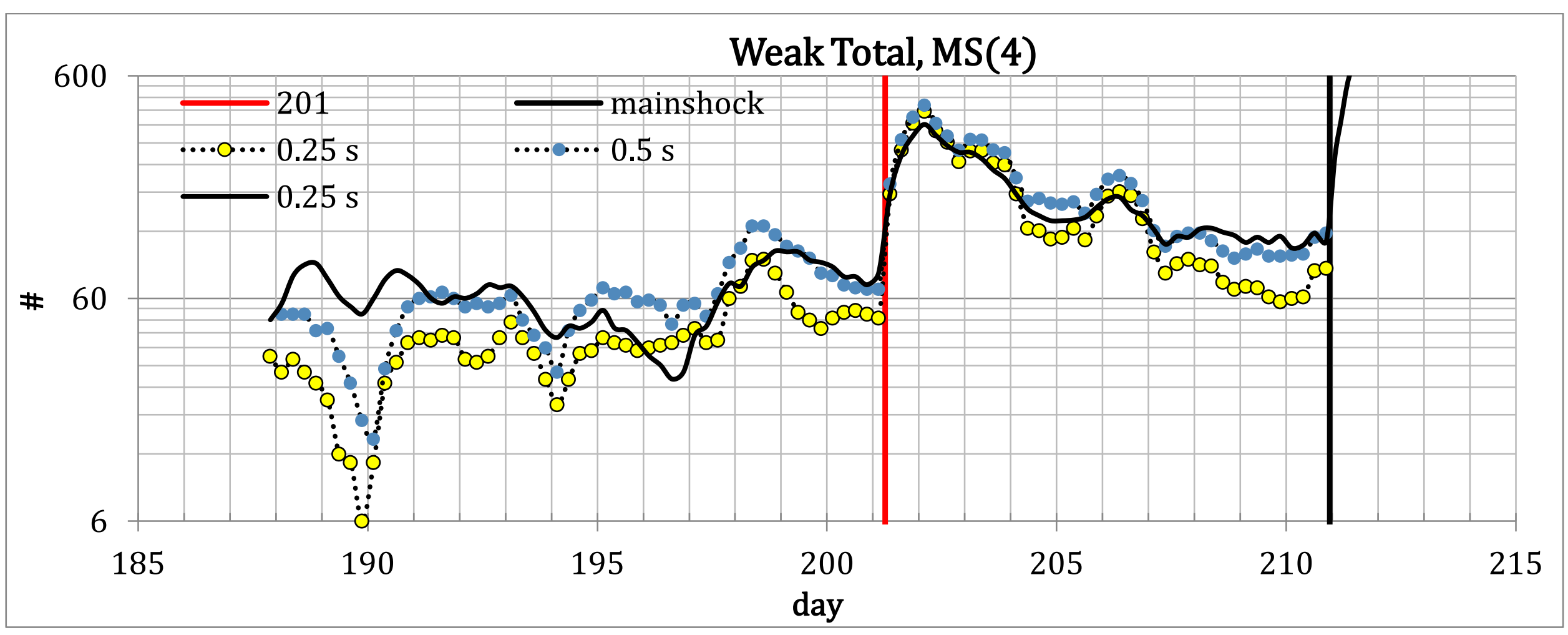

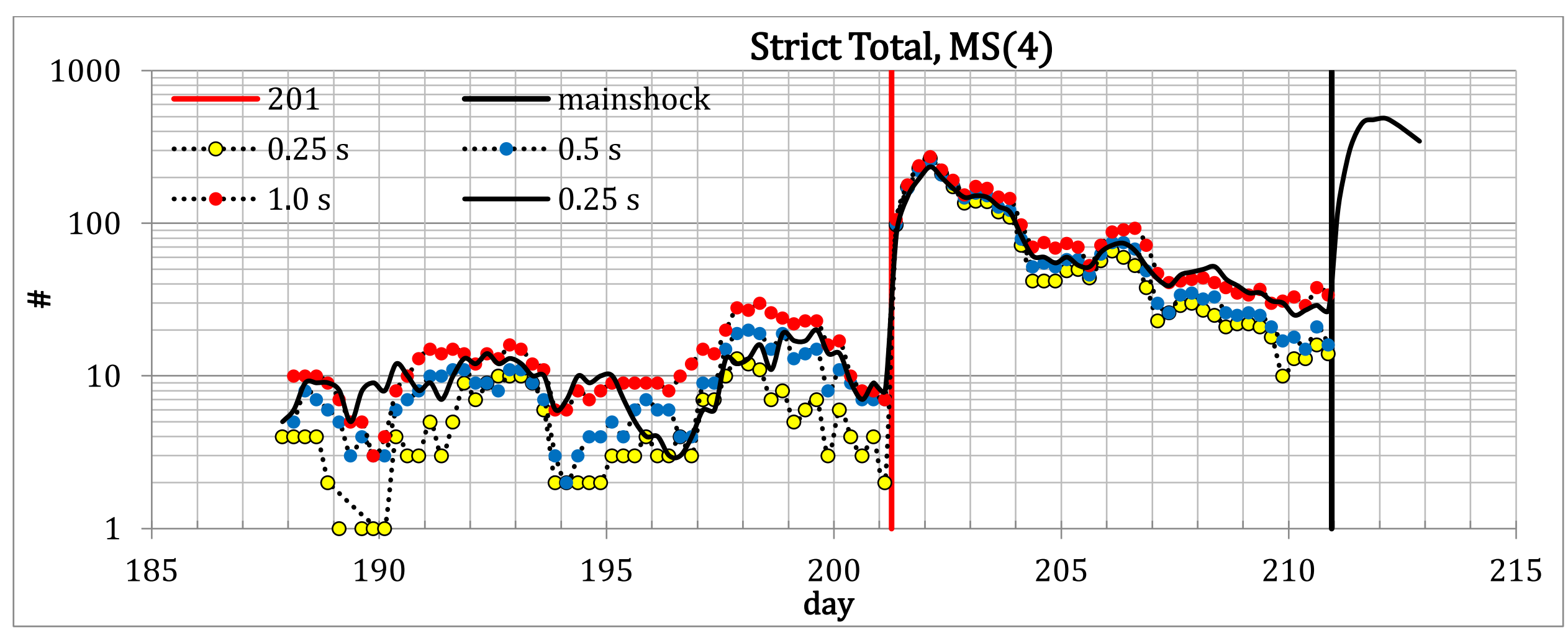

Figure 8. Comparison of the *StN*=0.0 and 0.075 curves. Thick black line represents the *StN*=0.075 case. Upper panel: the weak LA version. Lower panel: the strict LA version.

The result of the shift in the corner magnitudes of the twelve EDC sets associated with the enhanced WCC processing in the current study can be presented as a continuation of the REB recurrence curve in the A-zone into the low-magnitude range. The XSEL recurrence curve in Figure 9 for the 201-210 time period is calculated by conversion of the twelve indices into a linear set of magnitude thresholds. The $EDC_{12}$ is below the corner magnitude of the REB recurrence curve. The linear relationship changes the slope of the XSEL curve to that observed in the REB (-1.92) and thus converts integer numbers from 1 to 12 into a sequence of $m_b$ values from 3.82 to 3.19. The XSEL recurrence curve is exponential and there is no sign of its corner magnitude. The change in the EDC definitions may extend the XSEL curve in the low-magnitude direction. A similar effect can be achieved with the enhancement of the WCC processing by a range of *StN* values allowing for accurate tuning to local noise and maximum-amplitude conditions.

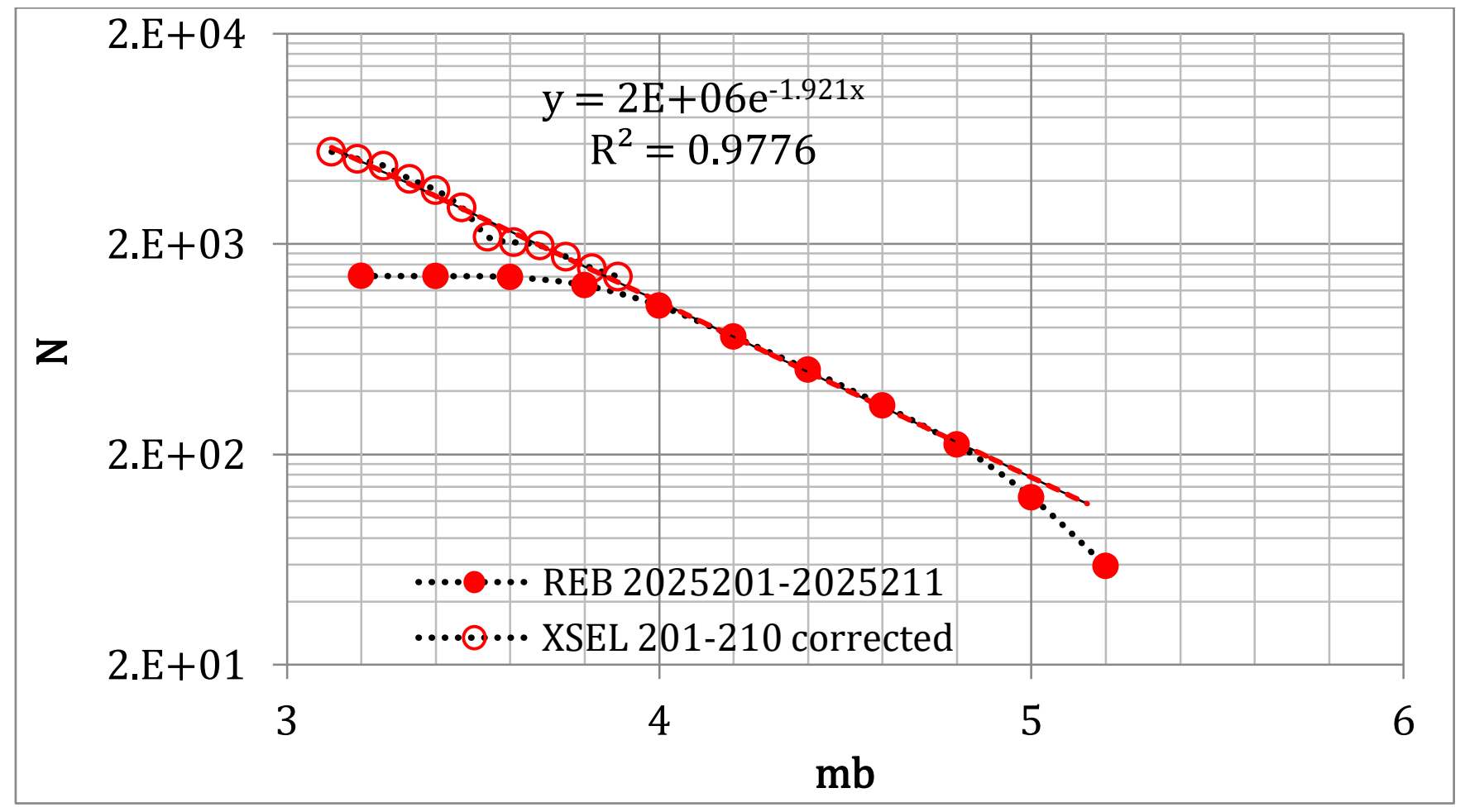

Figure 9. Recurrence curves as obtained from the REB and the XSEL for the period between 201 and 210 with intense aftershock sequence of the J20 event

The *StN* value of 0.075 was tuned to a specific time interval in [Kitov, 2026e] and may not be the best choice for whole processed period. Figure 10 presents the XSEL curves for the 2.0 s origin time tolerance and two cases with *StN*=0.0 and 0.075. The same set of the best 100

MEs is used and the XSEL are calculated for the A-zone. The original setting is more effective during the periods of intense seismicity. The *StN*=0.075 setting is more productive in terms of the number of XSEL events before the J20 and J29 events. For the origin time tolerances of 0.25 s and 0.5s this increase is important to achieve larger statistical significance of the weak/strict LA version ratio used as an indicator of approaching major rupture. In the original study, there were just a few XSEL events in some 24-h intervals. Consequently, any change by one or two XSEL events in the strict LA version leads to significant variation in this ratio.

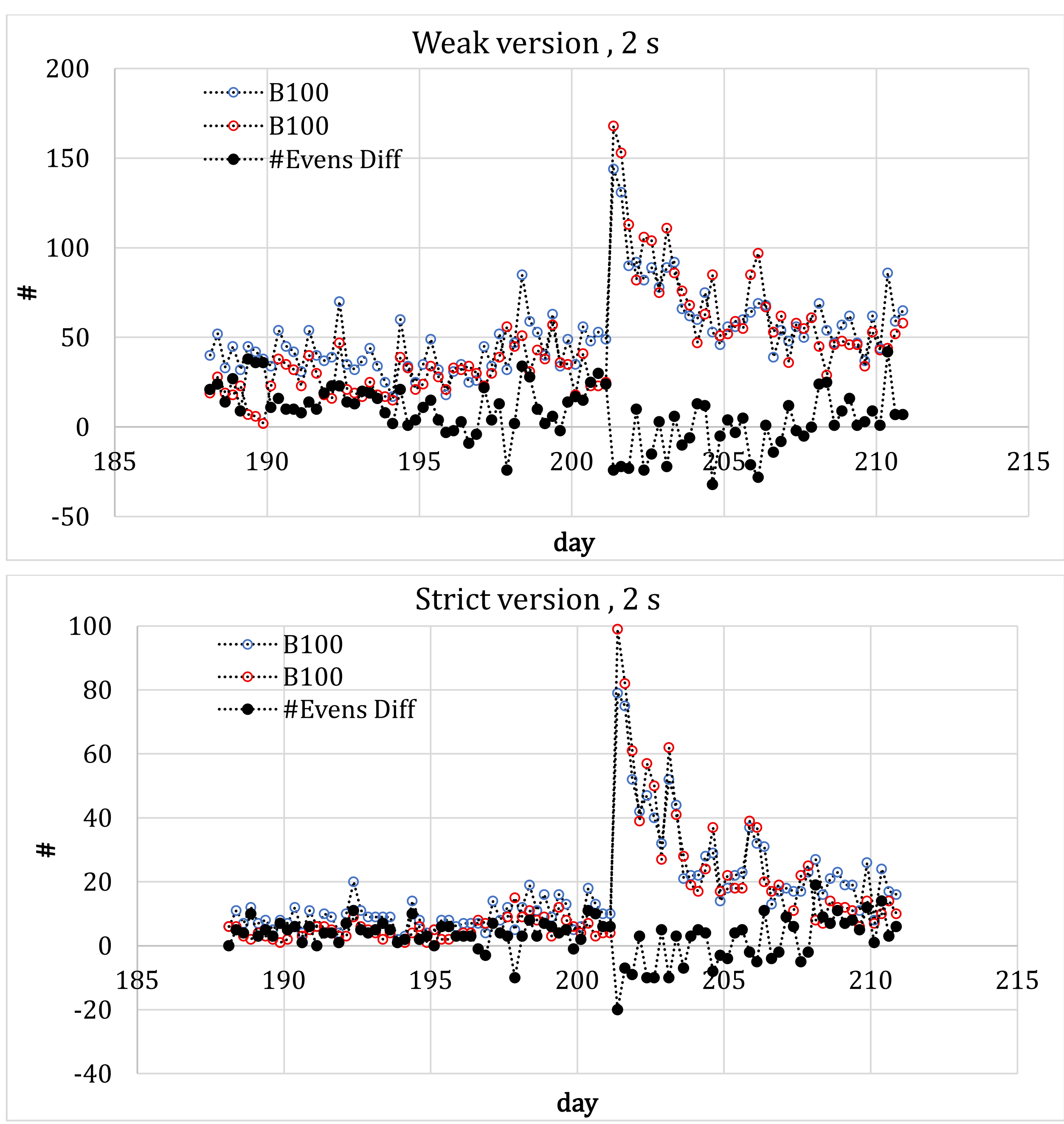

Figure 10. Relative performance of the 100 best MEs within the A-zone for *StN*=0.0 (original configuration without stochastic noise) and *StN*=0.075. Upper panel: the weak LA version with 2.0 s origin time tolerance. Lower panel: the strict LA version with the same tolerance.

Overall, the case with random noise is superior, but there were also periods of regular seismicity before the J20 earthquake when the addition of random noise had a negative effect on the XSELs for both the weak and the strict LA versions. Selected statistical estimates of the relative performance of the two *StN* cases are presented in Table 1. The numbers of XSEL events obtained for the three time periods are compared for the *StN*=0.0 and 0.075. The ratio of the respective XSELs illustrates the gain of the stochastic noise addition. The origin time tolerance of 0.25 s for the strict LA version is characterized by XSELs ratio of 1.97 for the period between days 187 and 201, and 1.13 for the period between days 201 and 210. For the whole period between days 187 and 210, the ratio is 1.21. For the largest tolerance of 5.0s the gain is negative for the period of high seismicity for both the strict and weak LA versions.

Table 1. Selected statistics for the ratio of the XSELs for *StN*=0.075 and 0.0

| Asperity | | Strict | | | | | Weak | | | |
|---|---|---|---|---|---|---|---|---|---|---|
| OT tolerance | days | StN=0.075 | StN=0.0 | Diff | "0.075"/"0.0" | | 0.075 | 0.0 | Diff | "0.075"/"0.0" |
| 0.25 s | 187-201 | 136 | 69 | 67 | 1.97 | | 892 | 562 | 330 | 1.59 |
| 0.25 s | 187-210 | 914 | 757 | 157 | 1.21 | | 2550 | 2091 | 459 | 1.22 |
| 0.25 s | 201-210 | 778 | 688 | 90 | 1.13 | | 1658 | 1529 | 129 | 1.08 |
| 1.0 s | 187-201 | 346 | 178 | 168 | 1.94 | | 1770 | 1157 | 613 | 1.53 |
| 1.0 s | 187-210 | 1313 | 1071 | 242 | 1.23 | | 4036 | 3394 | 642 | 1.19 |
| 1.0 s | 201-210 | 967 | 893 | 74 | 1.08 | | 2266 | 2237 | 29 | 1.01 |
| 2.0 s | 187-201 | 512 | 285 | 227 | 1.80 | | 2358 | 1616 | 742 | 1.46 |
| 2.0 s | 187-210 | 1606 | 1327 | 279 | 1.21 | | 4977 | 4288 | 689 | 1.16 |
| 2.0 s | 201-210 | 1094 | 1042 | 52 | 1.05 | | 2619 | 2672 | -53 | 0.98 |
| 5.0 s | 187-201 | 766 | 485 | 281 | 1.58 | | 3237 | 2518 | 719 | 1.29 |
| 5.0 s | 187-210 | 1969 | 1780 | 189 | 1.11 | | 6288 | 5876 | 412 | 1.07 |
| 5.0 s | 201-210 | 1203 | 1295 | -92 | 0.93 | | 3051 | 3358 | -307 | 0.91 |

The effect of stochastic noise addition to the waveforms before calculation of the *CC* in the WCC pipeline is aimed at the improvement in the estimates of the predictive parameters associated with the evolution of low-magnitude seismicity from the lowest detectable values to the level immediately before the coseismic phase. The magnitudes of the events occurring days and hours before the J20 earthquake are nor accurately measure in the ambient noise, but they are definitely below the detection threshold of the REB as Figure 3 demonstrates. The initiation of the J29 earthquake is masked by the aftershocks sequence of the J20 event. It is difficult to distinguish these aftershocks and the foreshocks of the J29 event in the REB. Therefore, the evolution of low-magnitude events is likely the way to measure earthquake preparation effects [Schaff *et al.*, 2025] and to obtain quantitative parameters describing this evolution.

In the original paper [Kitov, 2026d], the XSEL was obtained for various sets of MEs, including the best 100. In the current study, we use a LA scheme developed in [Kitov, 2026e], which allows for the improvement of the approach developed in the original study. The MEs in the O- zone are not excluded from the WCC processing as they can suppress the side-sensitivity of the array stations used in the calculations. This effect allows for the avoidance the study-irrelevant XSEL events generated by the MEs far beyond the preparation zone.

Table 2 presents selected statistics of the 100 best MEs in the A-zone and O-zone. This is an important separation when the target area is within a larger area with MEs. The ratio of the numbers of XSEL events in the A-zone and O-zone (O/A) is less than 1.0 for all cases in line with the 63:37 proportion of the MEs. The lowest O-zone relative activity is observed during the periods of the highest aftershock activity in the A-zone. For the period of regular activity from day 187 to 201, the O-zone activity is relatively high and, if not excluded, is able to introduce significant bias in the estimates of the ratios of the weak and strict LA versions for the same origin time tolerance. The effect of the side-sensitivity of the WCC detector at array stations may not be too large on average, but its local variations can introduce substantial bias into the predictive parameters, which must be suppressed.

Table 2. Selected statistics for the ratio of the XSELs in A-zone and O-zone for the *StN*=0.075 case.

| | | Strict XSEL | | | | | Weak XSEL | | | |
|---|---|---|---|---|---|---|---|---|---|---|
| OT tolerance | days | All | A-zone | O-zone | O/A | | All | A-zone | O-zone | O/A |
| 0.25 s | 187-201 | 227 | 136 | 91 | 0.67 | | 1609 | 892 | 717 | 0.80 |
| 0.25 s | 187-210 | 1148 | 914 | 234 | 0.26 | | 4133 | 2550 | 1583 | 0.62 |
| 0.25 s | 201-210 | 921 | 778 | 143 | 0.18 | | 2524 | 1658 | 866 | 0.52 |
| 1.0 s | 187-201 | 580 | 346 | 234 | 0.68 | | 2953 | 1770 | 1183 | 0.67 |
| 1.0 s | 187-210 | 1884 | 1313 | 571 | 0.43 | | 6683 | 4036 | 2647 | 0.66 |
| 1.0 s | 201-210 | 1304 | 967 | 337 | 0.35 | | 3730 | 2266 | 1464 | 0.65 |
| 2.0 s | 187-201 | 868 | 512 | 356 | 0.70 | | 3866 | 2358 | 1508 | 0.64 |
| 2.0 s | 187-210 | 2396 | 1606 | 790 | 0.49 | | 8264 | 4977 | 3287 | 0.66 |
| 2.0 s | 201-210 | 1528 | 1094 | 434 | 0.40 | | 4398 | 2619 | 1779 | 0.68 |
| 5.0 s | 187-201 | 1237 | 766 | 471 | 0.61 | | 5323 | 3237 | 2086 | 0.64 |
| 5.0 s | 187-210 | 3094 | 1969 | 1125 | 0.57 | | 10595 | 6288 | 4307 | 0.68 |
| 5.0 s | 201-210 | 1857 | 1203 | 654 | 0.54 | | 5272 | 3051 | 2221 | 0.73 |

***Predictive parameters***

It is worth comparing the results of the original and the current WCC processing with the XSELs without separation into the A-zone and O-zone. Figures 11 and 12 show the weak-to-strict LA version ratios for the periods 187-201and 201-210, respectively. Such a comparison can reveal the differences in the WCC performance related to the capability of the stochastic noise addition to increase the SNRcc values of the originally detected signals and to detect new signals hidden below the detection threshold in the original approach. The period of regular, relatively low seismicity between days 187 and 201 in Figure 11 illustrates such differences near the origin time ("195.901") of an REB event with coordinates 46.66°N, 151.30°E, $m_b$(IDC)=5.4, and a depth of 87 km (see Figure 3). This event occurred in the O-zone and was likely not associated with the J20 earthquake preparation. The original curves for 0.25 and 0.5 s tolerances fall to the level of ~6 together with the other four curves. The same curves in the current study do not demonstrate such a fall.

The time interval between day 198 and 199 exhibits a sharp peak in the 0.25 s original curve, not observed in the current study. The period immediately prior to the J20 earthquake is characterized by the rise in three curves (0.25, 0.5, and 1.0 s) in the current study. Only two curves with lower tolerance demonstrate a clear increase in the original study. Overall, the patterns during this period are similar, and the surge in some curves during the two days prior to the J20 event can be considered as a very specific feature likely associated with the preparation process for this earthquake.

In Figure 12, the same two cases are shown for the period between days 201 and 210. The principal difference is observed during the day 210 prior to the J29 earthquake. The original curves for tolerances 0.25 s and 0.5 s increase in amplitude from 4.5-5.0 to ~7.0. The 1.0 s curve rises from 4.3 to a level of 5.4. The other three curves demonstrate weak positive trends that are not statistically significant. For the current study, only the 0.25 s curve demonstrates an increasing trend prior to the J29 earthquake. The other five curves do have visible positive trends but these trends are smoothed and suppressed by the increased number of the XSEL events in the WCC processing with stochastic noise.

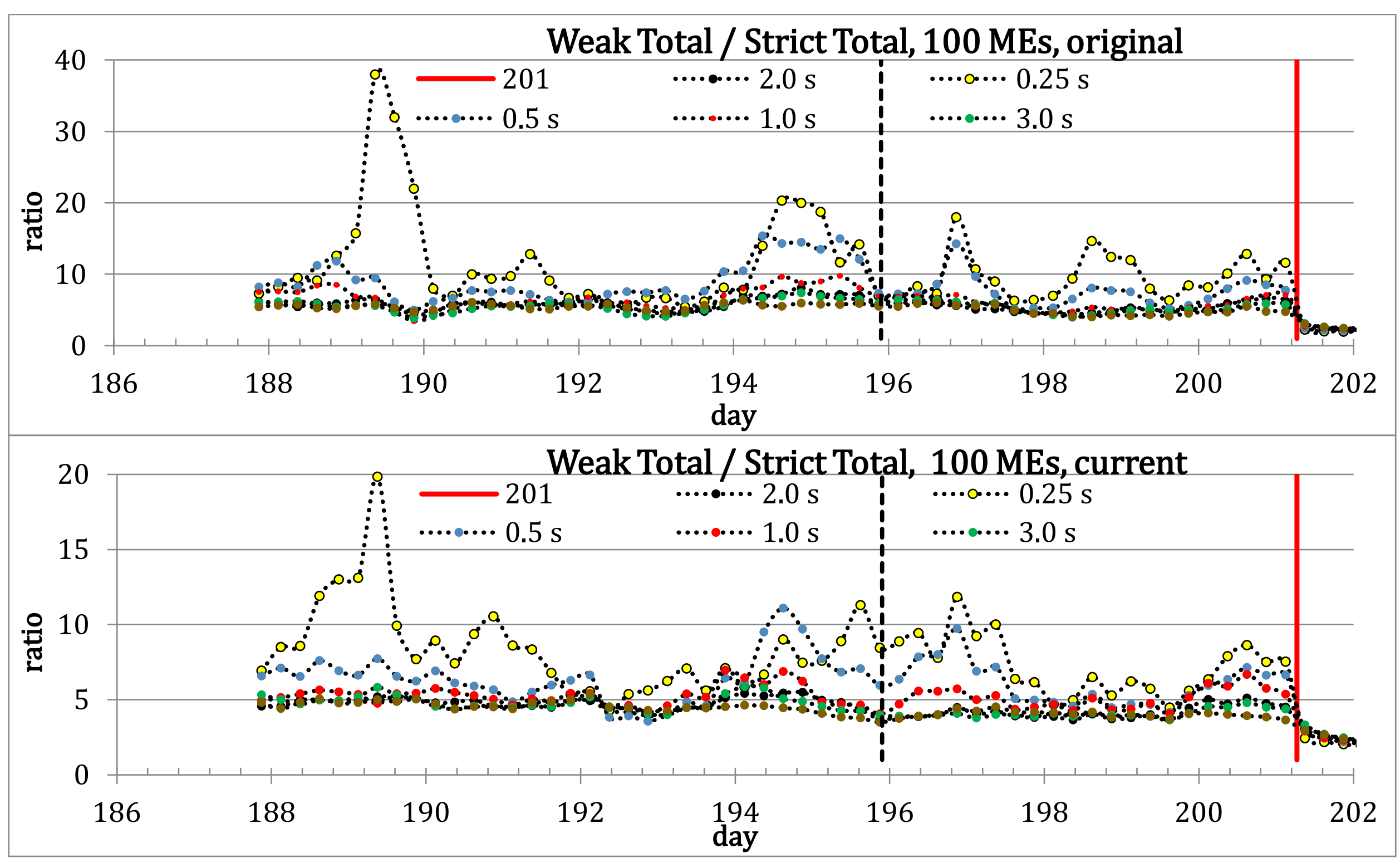

Figure 11. Comparison of the original and the current studies for the period 187-201. Total XSELs for the best 100 MEs are presented.

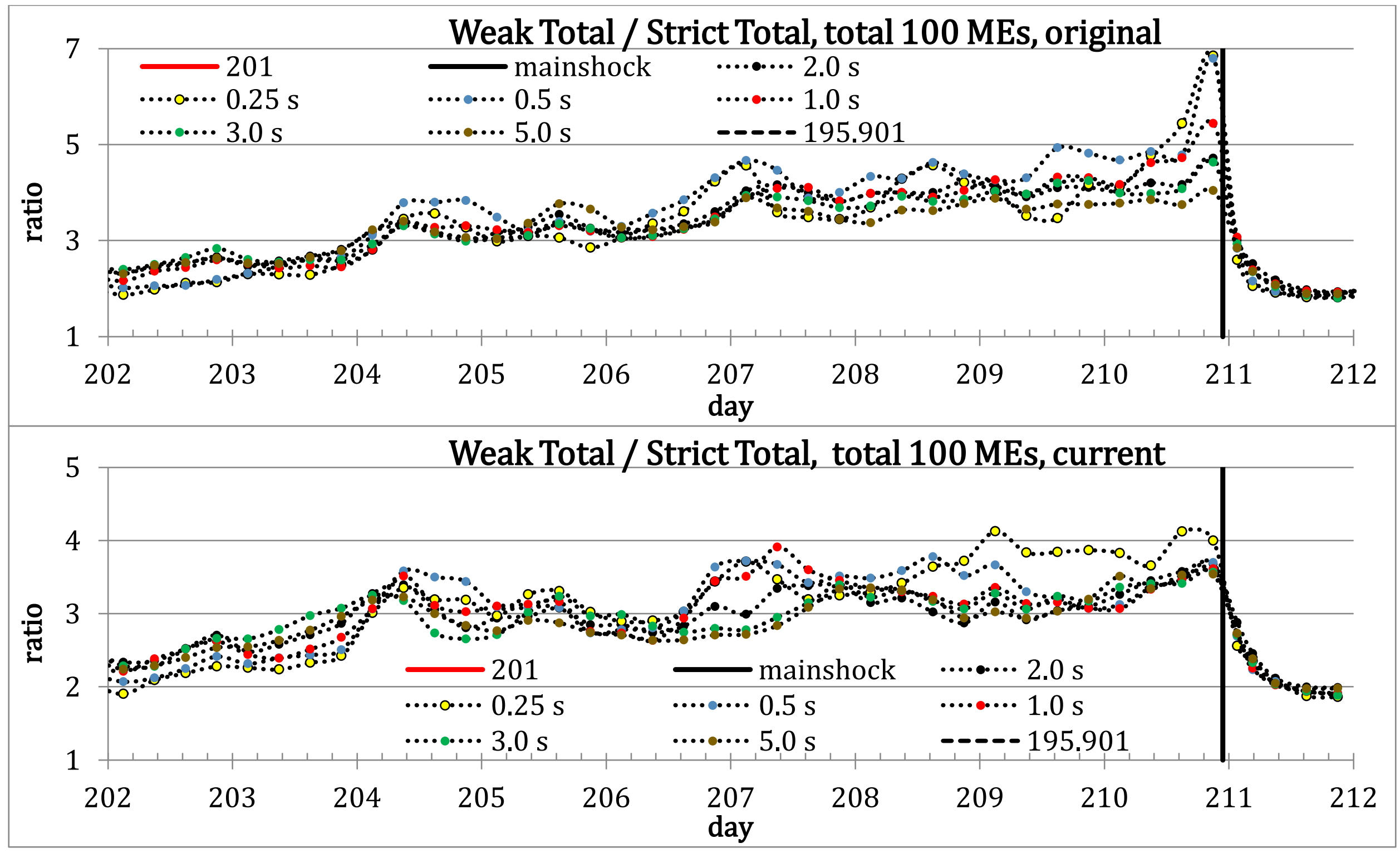

Figure 12. Comparison of the original and the current studies for the period 201-210. Total XSELs for the best 100 MEs are presented.

The predictive parameters for the A-zone are presented in Figure 13 for the days 187-201. In addition, the weak-to-strict LA ratios for the O-zone are shown for contrast. In the upper panel of Figure 13 the curves obtained for the original *StN*=0.0 setting, but corrected for the A-zone restriction for the XSEL events' location. The 0.25 s curve demonstrates a three-day-long oscillating increase with the peak of 25 in the last six hours before the J20 earthquake initiation. There is a three-day-long period between 195 and 197 (approximately around the 195.901 earthquake) when the 0.25 s and 0.5 s curves were elevated to the level of 15. A still period between these two high amplitude periods was one-day-long and all six curves were below the level of 5.

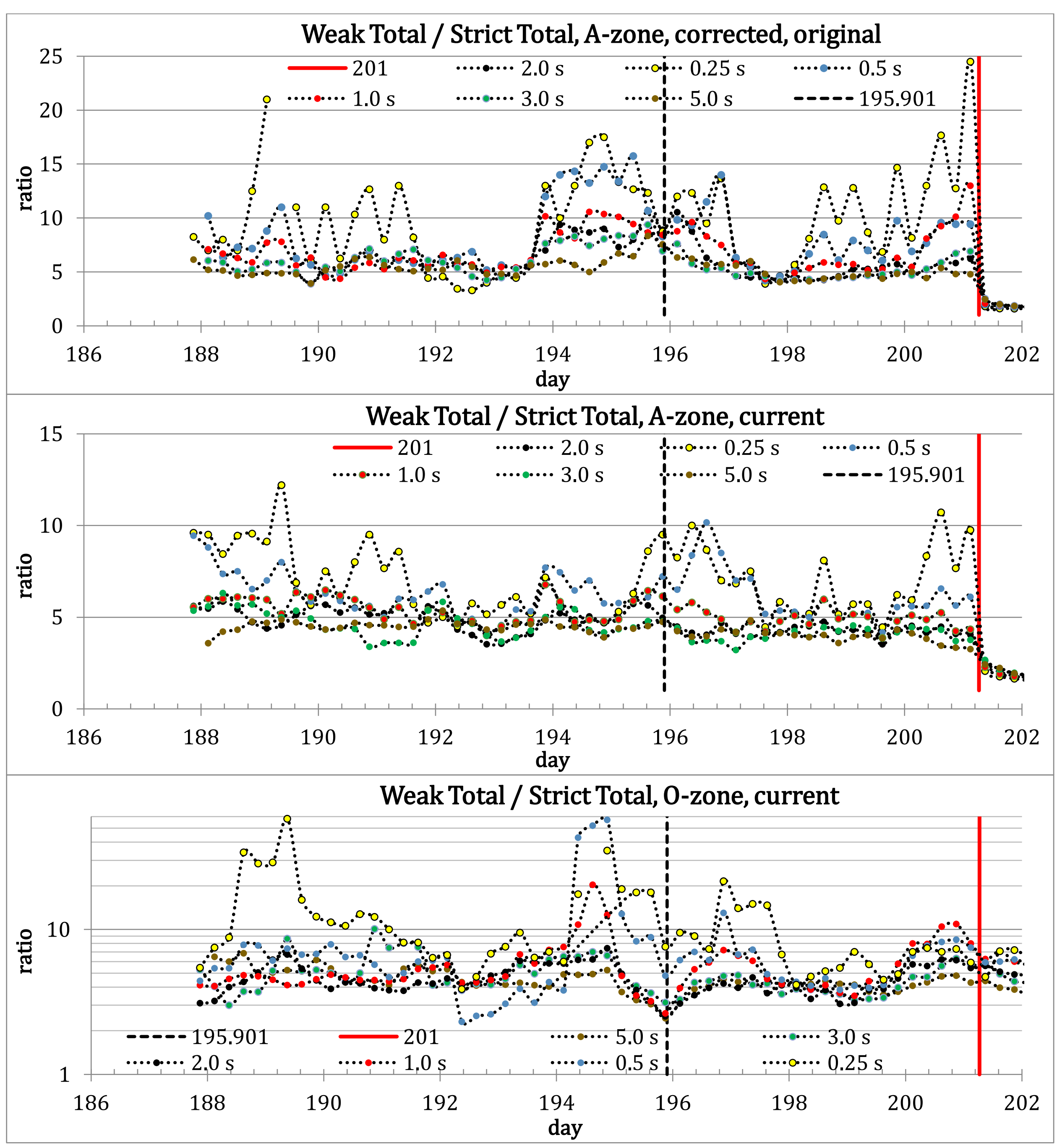

Figure 13. Evolution of the ratios of the numbers of XSEL events in the weak and the strict LA versions for six origin time tolerances between July 6 and July 21, 2025. Upper panel: the original study corrected for the separation of the A-zone and O-zone. Middle panel: the same for the current study. Lower panel: the curves for the O-zone in the current study.

In the middle panel of Figure 13, the results of the current study are shown. All six curves are lower than for the original case, likely due to higher numbers in the corresponding XSELs with the larger gain obtained by the strict LA version (see Figure 8). The three-day-long increase in the 0.25 s curved amplitude for the original case is now reduced to a one-day-long period of the curve elevated from approximately 5.0 to a level above 10.0. Other oscillations are suppressed by a better statistical coverage of the enhanced WCC processing. This peak prior to the J20 event could be a good predictive indicator, but the same surge was observed prior to the "195.901" event. The difference between these two cases is in their statistical significance. The strict LA version of the 0.25 s curve was twice as large for the pre-J20 case (see Figure 6).

The O-zone curves obtained as the difference between the total XSEL and that of the A-zone are shown in the lower panel of Figure 13. They are extremely volatile with the peaks of 60 observed in the 0.25 s and 0.5 s curves. The period prior to the J20 event shows the 1.0 s curve increasing to the level of 11.0 from ~4.0. The 0.5 s curve is the second highest, with the 0.25 s curve not demonstrating any significant increase. Such peaks are also observed during the whole period and are not useful for earthquake prediction.

The evolution of the predictive parameters —that is, the weak-to-strict XSEL ratio for six different origin-time tolerances, for the WCC processing with the random noise addition is the ultimate goal of the current study. Figure 14 is similar to Figure 13 but for the days between days 202-212. The upper panel presents the result of the original study but calculated for the A-zone XSEL events only. The 0.5 s curve exhibits the highest peak of ~7.4 in the last six hours prior to the July 29, 2025, Kamchatka megathrust earthquake. This curve started its increase to the peak 30-h before the J29 event from the level of ~4.0 with the total increase of ~85%. The 0.25 s curve is the second in amplitude prior to the J29 coseismic phase. The other four curves have a marginal growth or a slight fall. The 0.5 s and 0.25 s curves are significantly higher than any other peak before the second half of July 28 (day 209). They can be associated with the increasing seismicity in the earthquake preparation process and serve as predictive parameters.

The results of the current study presented in the middle panel are similar to those in the upper panel with the 0.5 s curve demonstrating the highest peak several hours prior to the J29 coseismic phase. This peak is ~ 4.9 with the start point at 2.8—that is, the curve has also increased by 75% in the last 30 hours before the mainshock. The 0.25 s and 1.0 s curves demonstrate approximately the same growth except the former started slightly earlier to increase to its final peak of 4.0 from 2.4 in the second half of day 208. The 1.0 s curve started to increase one day later.

The most significant changes to the weak-to-strict LA versions' ratios are observed in the curves with larger origin time tolerances from 2.0 s to 5.0 s. They all demonstrate significant growth from the level of 2.9 to 3.9 during the 30-h period prior to the mainshock. Figure 6 suggests that this growth is related to the increasing number of XSEL events for all curves of the weak LA version, with the strict XSEL demonstrating almost constant numbers for the same period. This could be the effect of higher WCC resolution in the magnitude range associated with the weak LA version. These events have been missed by the standard WCC processing without random noise.

The lower panel with the O-zone results supports the assumption of the preparation process concentrated within the A-zone. The highest peak of 13.1 in the 0.25 s curve was observed on July 23. There was a J20 aftershock with coordinates 52.55°N, 160.51°E and $m_b$(IDC)=5.85 on July 22, close to the epicenters of the J20 and J29 earthquakes. Therefore, the peak on July 23 is likely not the cause of the surge in the 0.25 s curve in the O-zone. There are no indications of a change in the seismic process similar to that observed in the A-zone as identified by the WCC processing days prior to the J29 earthquake. All six curves gradually converge to the level of 2.6-4.2 more than a day before the mainshock. Interestingly, this convergence is driven by the increasing number of events in the strict XSELs as shown in Figure 15. A similar effect was observed two days prior to the J20 event (Figure 13), where all the curves except that for the 0.25 s tolerance were within the 3.5-4.5 range. This leveling effect lasted approximately one day and then transformed into an increasing trend, which peaked approximately 12 hours before the J20 event. These observations can be associated with the interaction between the A-zone and O-zone seismicity and need further investigation.

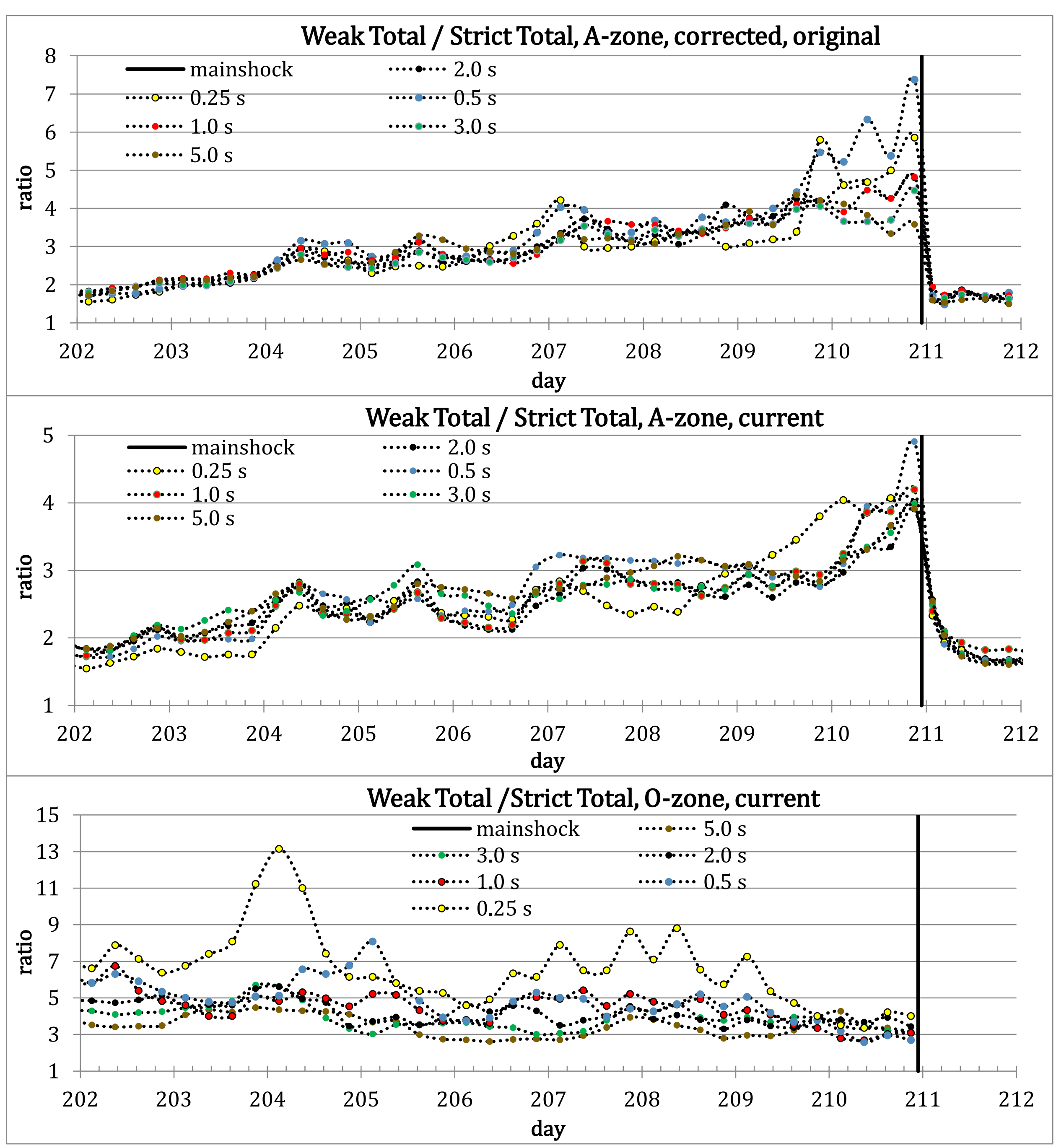

Figure 14. Same as in Figure 13 for the period from July 21 to July 30, 2025.

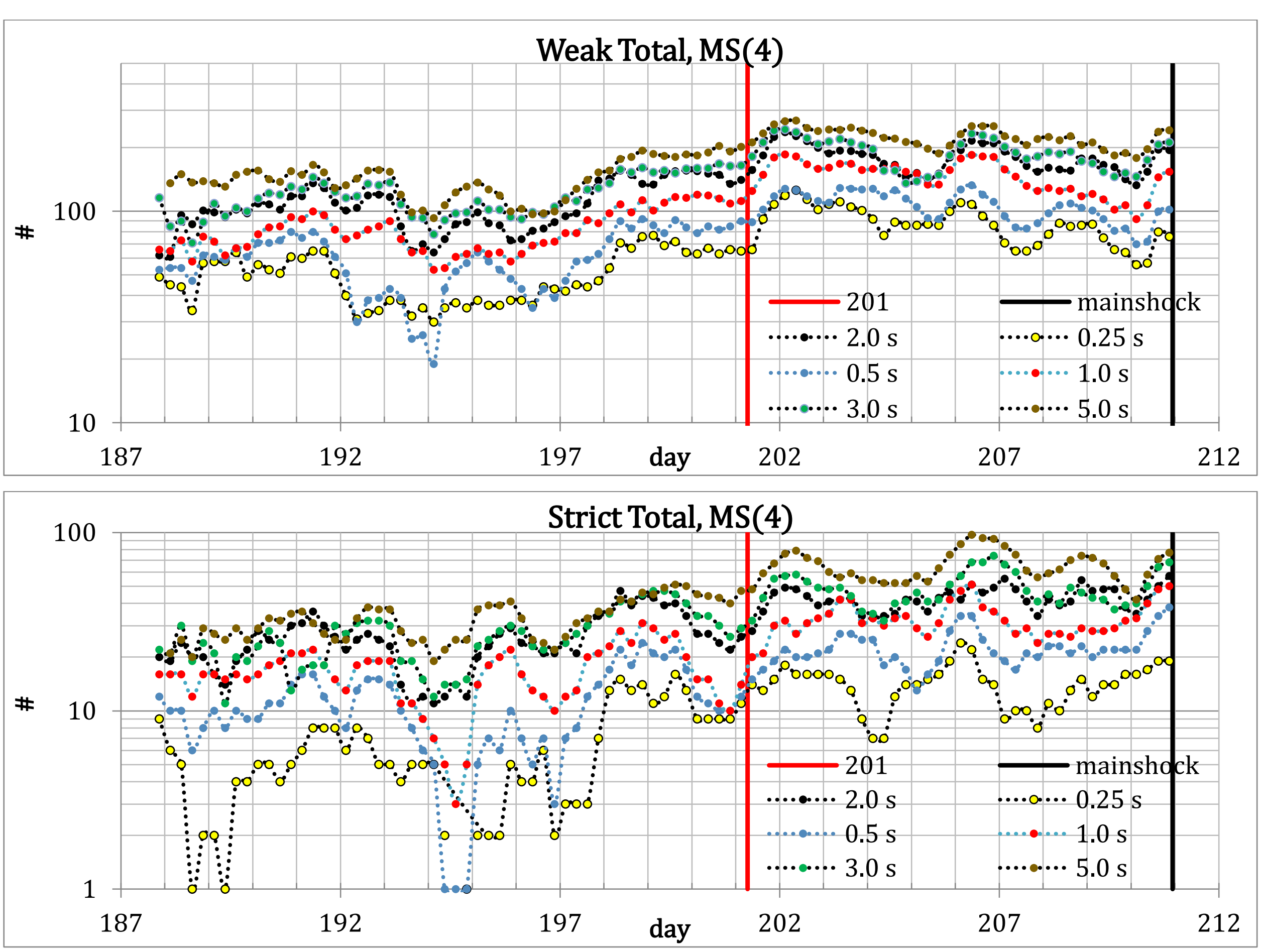

Figure 15. Evolution of the number of events in XSELs for the weak (upper panel) and strict (lower panel) LA versions for six origin time tolerances.

## Discussion

The study of earthquake preparation processes is important as a civilian application and also as an opportunity to understand physical processes in the Earth. The first objective suggests a reliable forecast of future mega-earthquakes using statistically significant, measurable parameters. Physical processes prior to the largest earthquakes can be studied retrospectively. This study is based on the increasing capabilities of the global and regional observational systems, including the seismic network of the International Monitoring System. The exceptional sensitivity and resolution of the IMS arrays can be substantially improved by a denser set of 3-C stations of the IMS and other geophysical agencies and institutions. These stations have to meet the strict quality requirements, including a low ambient noise level, to be able to detect ultra-weak signals from low-magnitude events of interest for earthquake prediction.

High-quality datasets are most appropriate for enhanced methods of processing, allowing for significant improvements in the signal detection and phase association. Waveform cross-correlation is the core algorithm in the matched filter method, which is practically an almost optimal detector for repeating signals in the long-term cyclic evolution of the global tectonics with supercritical stresses mainly released via catastrophic earthquakes. The density and quality of potential master events have been increasing over time since the start of global seismological observations. The elevated level of seismicity has also a negative effect on seismic observations: the level of ambient noise coherent with the sought signals is also elevated and reduces the detection capabilities of the matched filter detector, which requires random noise for an optimal performance.

The principal result of the current study is the synchronous increase by a factor of 1.5 to 2.0 in all six curves prior to the July 29, 2025, Kamchatka earthquake. It is a rare observation before a catastrophic earthquake. It can be considered as strong evidence in favor of the possibility of measuring the processes of earthquake preparation expressed in the growing seismic activity during tens of hours prior to the start of the coseismic phase. The WCC processing with added stochastic noise has reached this goal with one sub-optimal *StN* value.

Further improvements are possible in this direction. A set of several *StN* values can be tuned to fit a broader range of maximum amplitudes in the processing intervals. These peak amplitudes can be associated with larger sources at teleseismic distances or regional sources with lower magnitudes which affect the regional network most sensitive to the sought signals. Local measures of amplitude, such as STA and LTA can be used instead of the global parameter of the maximum amplitude. The corresponding scaling factors for these two measures should also be adjustable and cover an appropriate range.

The geometrical configuration of the earthquake preparation zone is not known in advance. The retrospective assumption that the aftershock zone is a good approximation is not fully correct, since the epicenters of the J20 and J29 earthquakes were within tens of kilometers of each other. The epicenter of the May 19, 2013, Kamchatka M6.0 earthquake was also very close to the J20 and J29 epicenters. We can focus the WCC processing on the smaller area shown in Figure 1 by a yellow oval. There is a problem with the number of XSEL event needed to make the ratios of weak-to-strict LA versions statistically significant. The smaller the studied area, the lower the number of XSEL events. Therefore, the choice of the preparation zone has to be flexible. Since the XSEL is obtained in the LA and CR processes, when the final WCC detection list is available, multiple configurations of the A-zone can be processed and compared. This provides substantial flexibility in the WCC processing.